\documentclass[aps, prx, superscriptaddress, twocolumn, letterpaper, 10pt, amsfonts, amsmath, amssymb]{revtex4-2}

\usepackage{graphicx}
\usepackage{hyperref} 
\usepackage{xcolor}
\usepackage{soul}
\hypersetup{
    colorlinks,
    linkcolor={blue!80!black},
    citecolor={blue!80!black},
    urlcolor={blue!80!black}
}
\usepackage{physics}
\usepackage{braket}
\usepackage{comment}

\usepackage{placeins}

\usepackage{siunitx}
\usepackage{chemformula}

\usepackage{xcolor}
\usepackage{lipsum}
\usepackage{booktabs}

\usepackage{titletoc}

\titlecontents{section}[0em]  
    {\vspace{0.5em}\bfseries} 
    {\thecontentslabel.\ }   
    {}
    {\titlerule*[0.5pc]{}\contentspage} 
    \titlecontents{subsection}[2em] 
    {\vspace{0.3em}}           
    {\thecontentslabel.\ }    
    {}
    {\titlerule*[0.5pc]{}\contentspage}

\definecolor{editcolor}{rgb}{1, 0.0, 0.0}

\newcommand{\equref}[1]{Eq.~(\ref{#1})}

\newcommand{\figref}[1]{Fig.~\ref{#1}}
\newcommand{\refcite}[1]{Ref.~\onlinecite{#1}} 
\renewcommand{\vec}[1]{\boldsymbol{#1}}

\begin{document}
\title{Altermagnetic superconducting diode effect \\ from non-collinear compensated magnetism in Mn$_3$Pt}

\author{Constantin Schrade}
\affiliation{Hearne Institute of Theoretical Physics, Department of Physics \& Astronomy, Louisiana State University, Baton Rouge LA 70803, USA}

\author{Sujit Manna}
\affiliation{Department of Physics, Indian Institute of Technology Delhi, Hauz Khas, New Delhi 110016, India}

\author{Mathias S. Scheurer}
\affiliation{Institute for Theoretical Physics III, University of Stuttgart, 70550 Stuttgart, Germany}

\date{\today}

\begin{abstract}
Altermagnets have recently emerged as a distinct class of magnetic systems that exhibit spin splitting of electronic bands while retaining zero net magnetization.
This unique combination makes them a promising platform for time-reversal symmetry-breaking superconducting phenomena,
although identifying concrete material platforms remains an important open challenge. Here, we develop a theory for the superconducting diode effect observed experimentally in a Mn$_3$Pt-superconductor heterostructure.  
Using both a symmetry analysis and model calculations on the breathing kagome lattice, we show how the altermagnetic spin textures in Mn$_3$Pt generate a spin splitting of the electronic bands that remains magnetization-free even in the presence of spin-orbit coupling and, upon taking into account the proximity coupling across the interface, produces a superconducting diode effect. 
We also demonstrate that the angular dependence of the critical current provides a probe of the magnetic order. 
We hope that our work will contribute to the understanding and further discovery of candidate materials for novel altermagnet-superconductor hybrid devices.
\end{abstract}

\maketitle

\section{Introduction}
Altermagnets \cite{PhysRevX.12.040501,Review3,magnetism5030017,PhysRevX.12.040002,RafaelReview} 
are a class of magnetic orders in which symmetry enforces a vanishing net magnetization, $\vec{M}=0$, 
while still permitting a finite spin splitting of the electronic bands, even in the absence of spin-orbit coupling. 
This combination sets altermagnets apart from ferromagnets with a nonzero magnetization, 
and from antiferromagnets, where $\vec{M}=0$ but the electronic bands remain spin degenerate.
Initial interest in altermagnets was motivated by their large intrinsic spin splittings and the resulting opportunities for spintronic applications. 
More recently, however, their broader potential has become apparent across several communities.
Examples include proposed realizations in cold-atom platforms \cite{PhysRevLett.132.263402}, theoretical studies of coupling to lattice degrees of freedom \cite{PhysRevB.108.144418,PhysRevB.109.144421,2025PhRvB.111m4429X}, and altermagnets as candidate systems for topologically ordered phases \cite{Sobral2025May,Neehus2025Apr,Vijayvargia2025Oct}.

\begin{figure}[!t]
    \centering
    \includegraphics[width=0.9\linewidth]{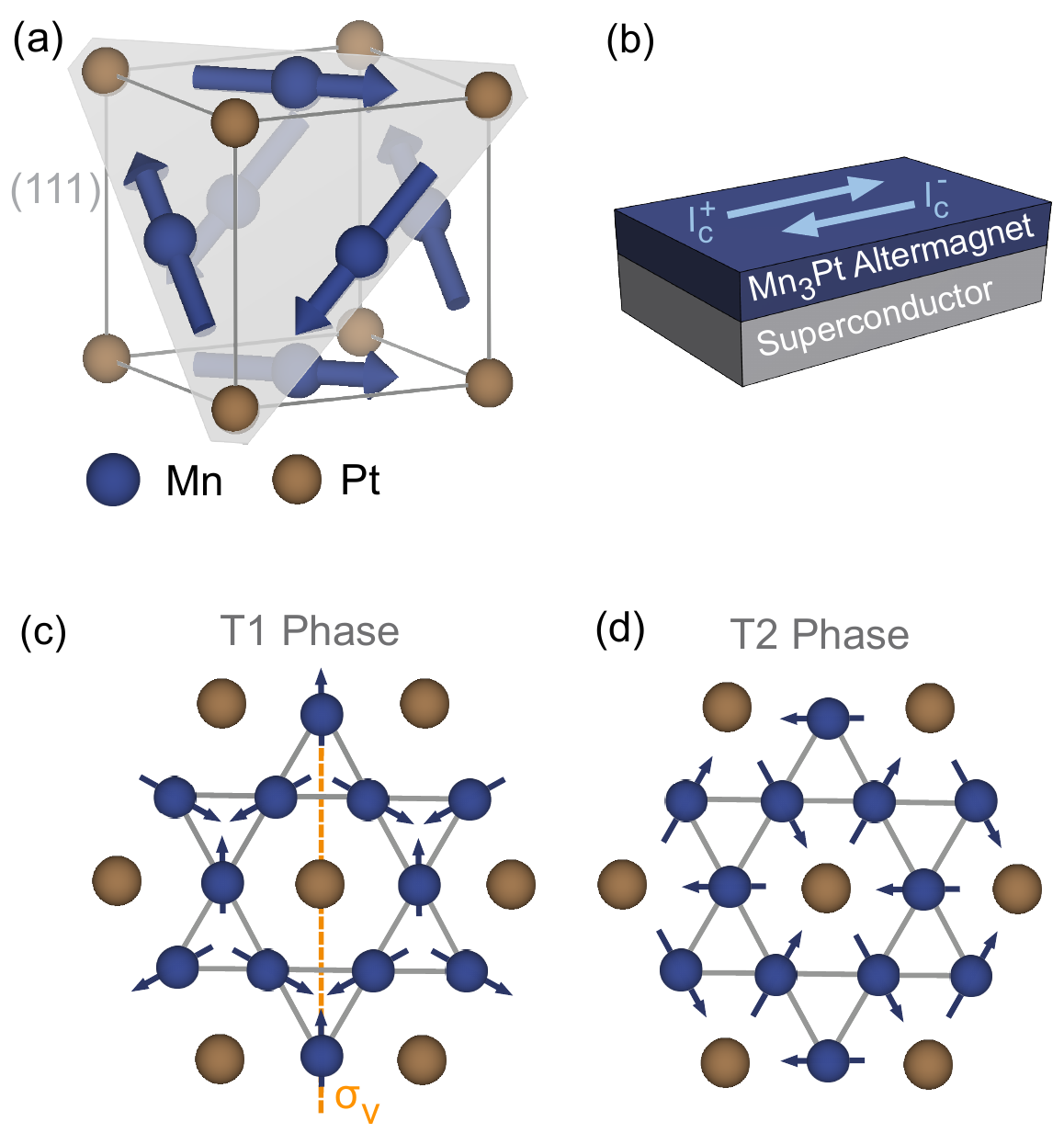}
    \caption{
\textbf{Noncollinear altermagnetic platform for a superconducting diode effect.}
(a) Crystal structure of Mn$_3$Pt, where Mn moments form a breathing kagome lattice in the $(111)$ plane (blue: Mn, brown: Pt). 
(b) Mn$_3$Pt altermagnet proximitized by a conventional superconductor, giving rise to a SDE 
with nonreciprocal critical currents, $I_c^+ \neq I_c^-$.
(c,d) Two compensated magnetic orders of Mn$_3$Pt: the \textsf{T1} phase (c) and the \textsf{T2} phase (d). The \textsf{T2} phase enables a SDE while retaining zero net magnetization. 
    }
    \label{fig:1}
\end{figure}

There is also a growing interest in the connections between altermagnetism and superconductivity \cite{Fukaya_2025,dlpb-gfct,dlpb-gfct,2025arXiv250909959M,PhysRevB.108.054510,2025arXiv250906790Z,2025arXiv251018102B,2025arXiv251026894A,2025arXiv250903774H,FiniteqPairing,PhysRevB.108.054510,PhysRevB.111.174436,PhysRevB.111.L100502,2025arXiv251019943S}. In particular, it was shown in \refcite{banerjee2024altermagnetic} that superconducting altermagnets 
can exhibit a superconducting diode effect (SDE) \cite{DiodeReview1,DiodeReview2} even at zero net magnetization. 
This means that the critical current $I_c(\hat{\vec{n}})$ of the superconductor along direction $\hat{\vec{n}}$ obeys $I_c(\hat{\vec{n}}) \neq I_c(-\hat{\vec{n}})$ for generic $\hat{\vec{n}}$, reflecting the breaking of time-reversal symmetry and any additional symmetries that would otherwise relate transport along $\hat{\vec{n}}$ and $-\hat{\vec{n}}$.
While several realizations of the SDE have been demonstrated in systems with $\vec{M}\neq 0$ \cite{Lin2022,FerromagneticProximity,le2024superconducting,wu2022fieldfree,scammell_theory_2022,zgnk-rw1p,chen2025finitemomentumsuperconductivitychiralbands,yoon2025quartermetalsuperconductivity,ingla2025,telkamp2025voltagetunablefieldfreejosephsondiode,PhysRevB.107.245415,PhysRevApplied.21.064029,35ss-brd7},
maintaining $\vec{M}=0$ imposes much stronger symmetry constraints on both the crystal point group and the altermagnetic order parameters \cite{banerjee2024altermagnetic}.
Despite the theoretical interest in altermagnetic SDEs \cite{b7rh-v7nq,2025PhRvL.135b6001C,yqsg-xdg8,2025arXiv250721446S,2025arXiv251201415R,2025arXiv251216260H} and magnetic Josephson junctions \cite{frazier2025spatiallyinhomogeneoustripletpairing,chaou2025proximitysuperconductivitychiralkagome,7j5m-g5xy,zhang2025finitemomentummixedsinglettripletpairing},
experimental progress on the impact of spin textures on superconductivity is still at an early but rapidly developing stage \cite{KAZMIN2025416602,Parkin1,Parkin2}.

In a very recent experiment\,\cite{PartnerExperimentalPaper}, the first experimental evidence of an altermagnetic SDE at zero external magnetic field was reported in a system of a Nb superconductor proximitized with the non-collinear magnet Mn$_3$Pt, see \figref{fig:1}(a,b).
Neglecting spin-orbit coupling for now, the magnetic order of Mn$_3$Pt \cite{kren1968magnetic,JKubler_1988, rimmler2023atomic, PhysRevB.92.144426,rimmler2025non,PhysRevB.92.144426} in the (111) interface plane of the heterostructure is shown in \figref{fig:1}(a). The combination of three-fold (spinful) rotational symmetry $C^s_{3z}$ perpendicular to the interface plane and the mirror plane $\sigma_v^s$ guarantee the vanishing of $\vec{M}$. At the same time, the magnetic texture at the interface breaks all symmetries that protect the spin degeneracy of the electronic bands and is expected to induce a finite, time-reversal-symmetry-breaking, spin splitting. As such, it can be thought of as altermagnetism \cite{NonColl2,NonColl3}, which was initially defined only with respect to collinear orders but has subsequently been extended \cite{NonColl3,PhysRevLett.132.176702,PhysRevMaterials.5.014409,PhysRevB.101.220403,NonColl1,NonColl2,PhysRevB.110.214428,SpinSplittingExp,PhysRevB.109.024404} as applicable to us here as well.  
Spin-orbit coupling breaks the continuous spin-rotation symmetry such that the spin texture shown in \figref{fig:1}(c) (referred to as $T_1$) and, e.g., the one rotated by $90^\circ$ ($T_2$ order \footnote{We here follow previous literature \cite{JKubler_1988} but note that this form of magnetic order is called \textsf{T3} in \cite{PhysRevB.92.144426}. Further, what we call \textsf{T3} is referred to as \textsf{T2} in \cite{PhysRevB.92.144426}.}), see \figref{fig:1}(d), become inequivalent, albeit energetically close \cite{PhysRevB.92.144426}. In \refcite{PartnerExperimentalPaper}, both orders are studied and both stabilize a SDE at zero external magnetic field. While the $T_1$ configuration is expected to give rise to a small canting and, hence, small net magnetization, $T_2$ realizes one of the few, idealized  scenarios of \refcite{banerjee2024altermagnetic} -- a SDE with symmetry-protected vanishing of the magnetization.

While this phenomenon was discussed in minimal single-band models and based on symmetry across different point groups in \cite{banerjee2024altermagnetic}, the material realization and findings in the experiment \cite{PartnerExperimentalPaper} pose several important open questions; these are related to the specific form of magnetic order in Mn$_3$Pt/Nb, its relation to altermagnetism, as well as the proximity effect across the heterostructure and will be addressed theoretically here. 
More specifically, we provide a theoretical analysis of the SDE in the Mn$_3$Pt/Nb heterostructure, taking into account the proximity coupling and the magnetic moments on the kagome lattice in Mn$_3$Pt. We start with an illustration of the spin splitting of the Fermi surfaces and discuss in what sense the magnetic order at the interface is related to altermagnetism. We show how the proximity coupling to Nb, which we model as a simple, reciprocal $s$-wave superconductor, induces critical current asymmetries in the latter. We also demonstrate how the directional dependence of the critical-current asymmetry could be used in future experiments to distinguish between the $T_1$ and $T_2$ magnetic orders and provide novel signatures of altermagnetic superconductivity.

\section{$\text{Mn}_{3}\text{Pt}$ and noncollinear altermagnetism}

\subsection{Noncollinear altermagnetism}
Although originally developed for collinear spin textures \cite{PhysRevX.12.040501}, the notion of altermagnetism has recently been extended to noncollinear spin textures as well \cite{NonColl3,PhysRevLett.132.176702,PhysRevMaterials.5.014409,PhysRevB.101.220403,NonColl1,NonColl2,PhysRevB.110.214428,SpinSplittingExp,PhysRevB.109.024404}. 
Since there is, to the best of our knowledge, still an on-going debate on the definition of non-collinear altermagnetism at the time of writing, we next introduce the conventions we use here.

Altermagnetism is defined in the limit without spin-orbit coupling, where the magnetic orders can be classified using the non-relativistic spin groups, with elements $[g_s||g_r]$ consisting of independent spin ($g_s$) and real-space ($g_r$) transformations. A magnetic texture $\vec{S}_j$ is referred to as altermagnetic if it obeys the following properties: (1) there is a symmetry that ensures that the magnetization $\vec{M}_c = \sum_j \vec{S}_j$ vanishes, and (2) it lifts the spin degeneracy of the electronic bands. Property (1) distinguishes altermagnets from ferromagnets, while (2) separates them from antiferromagnets with spin-degenerate bands. We note that neither property depends on collinearity and both apply equally to \textit{noncollinear} magnetic crystals. It is therefore natural to adopt (1) and (2) as the general defining properties of altermagnetism.

Some but not all authors add an additional requirement for altermagnetism (see, e.g., \cite{RafaelReview}). To define it, let us introduce the momentum-space spin textures, $\vec{s}_n(\vec{k}) = \braket{\psi_{n\vec{k}}|\vec{\sigma}|\psi_{n\vec{k}}}$, associated with the $n^{\text{th}}$ band and their Bloch states $\ket{\psi_{n\vec{k}}}$ in the magnetically ordered state; here $\vec{\sigma} = (\sigma_x,\sigma_y,\sigma_z)^T$ are the three Pauli matrices in spin space. In the presence of inversion symmetry (or $[E||C_{2z}]$ in a two-dimensional system), one can now further distinguish between magnetic textures that are even and odd under this symmetry, leading to $\vec{s}_n(\vec{k}) = \vec{s}_n(-\vec{k})$ and $\vec{s}_n(\vec{k}) = -\vec{s}_n(-\vec{k})$; such magnetic orders are then referred to as \textit{altermagnets} and \textit{antialtermagnets}, respectively. Importantly, if inversion (or $[E||C_{2z}]$) is broken, the two are in general expected to mix. In that case, we will follow \cite{NonColl2} and refer to these orders simply as altermagnets, unless stated otherwise. The crucial difference between the collinear and noncollinear scenario is that only the former has a $[C_{\infty}||E]$ symmetry and thus $\vec{s}_n(\vec{k}) = \pm \vec{e}_s$ pointing along or against a $\vec{k}$-independent quantization direction. In the noncollinear case, in contrast, $\vec{s}_n(\vec{k})$ generally span at least a plane.

We note that the requirement of a symmetry in property (1) means that the magnetization vanishes exactly and, thus, beyond the classical definition above. In particular, we also automatically have $\vec{M} = \sum_{n,\vec{k}} f_n(\boldsymbol{k}) \vec{s}_n(\vec{k}) = 0$, where $f_n(\boldsymbol{k})$ is the equilibrium occupation of the Bloch state $\ket{\psi_{n\boldsymbol{k}}}$. We finally point out that altermagnets (or antialtermagnets) can further be classified into states where $\vec{M} = 0$ still persists even when any symmetry-allowed spin-orbit coupling term is included or states where a finite magnetization is induced. We will discuss examples of both in this work.

\subsection{Crystal and Magnetic Structure of Mn$_3$Pt}
As a next step, we will introduce Mn$_3$Pt as a concrete example of a noncollinear altermagnet.
Mn$_3$Pt crystallizes in a cubic structure, with Pt atoms at the cube corners and Mn atoms at the face-centered positions \cite{kren1968magnetic,JKubler_1988, rimmler2023atomic, PhysRevB.92.144426}. 
Projecting the Mn sublattices onto $(111)$ planes yields two-dimensional kagome lattices of Mn sites, stacked in an $ABC$ sequence along the $[111]$ direction.
This stacking renders upward- and downward-pointing triangles inequivalent.
As a result, the six-fold $C_6$ rotation symmetry of an ideal kagome lattice is broken, leaving only a three-fold $C_3$ rotation symmetry. The Mn layers therefore realize ``breathing" kagome lattices and, together with the reflection symmetry $\sigma_v$, the point group is $C_{3v}$. 

Before illustrating it using a concrete tight-binding model, we will now explain based on symmetries why Mn$_3$Pt on these breathing kagome lattices satisfies the defining properties (1) and (2) of a noncollinear altermagnet: First, the magnetic moments in Mn$_3$Pt reside primarily on the Mn atoms. Well below the N\'eel temperature, $T_N \sim 475$~K, the Mn moments form a coplanar $120^\circ$ triangular state \cite{kren1968magnetic,JKubler_1988, rimmler2023atomic, PhysRevB.92.144426,rimmler2025non,PhysRevB.92.144426}, like the one shown in \figref{fig:1}(c). Focusing again first on the case without spin-orbit coupling, any global rotation of this texture is degenerate and symmetry-equivalent, like the one shown in \figref{fig:1}(d). To see that (1) is obeyed, we note that $C_{3z}^s := [C_{3z}||C_{3z}]$ leads to $M_{x,y}=0$ while $[C_{2\hat{s}}||\sigma_v]$, with $\hat{s}$ pointing along the direction of the spin on the mirror plane, imposes $M_{z}=0$. Furthermore, there is no $\vec{k}$-local anti-unitary symmetry that squares to $-1$ which follows by noting that the point group without magnetic order does not contain a unitary symmetry relating $\vec{k}$ and $-\vec{k}$ that could be combined with time-reversal $\Theta_s$ to form such a symmetry. In fact, even in the ``non-breathing" limit, where $C^s_{2z}\Theta_s$ is a symmetry, it satisfies $(C^s_{2z}\Theta_s)^{2}=+1$ and thus does \textit{not} enforce a Kramers' degeneracy.

As a result of the coplanar nature of the magnetic order, it is invariant under the product of $\Theta_s$ and the spin-only rotation $[C_{2z}||E]$. This implies that the bands, $E_n(\vec{k})$, and thus the energetic spin splittings are even in momentum, $E_n(\vec{k}) = E_n(-\vec{k})$. For the spin texture, it imposes the constraint $[\vec{s}_n(\vec{k})]_{x,y} = [\vec{s}_n(-\vec{k})]_{x,y}$ and $[\vec{s}_n(\vec{k})]_{z} = -[\vec{s}_n(-\vec{k})]_{z}$. In the non-breathing limit, the symmetry $[E||C_{2z}]$ further leads to $\vec{s}_n(\vec{k}) = \vec{s}_n(-\vec{k})$ such that $[\vec{s}_n(\vec{k})]_{z} = 0$. Taken together, one can think of the breathing of the kagome lattice as admixing an antialtermagnetic component, purely along the spin-$z$ direction, while the spin splitting in the $xy$-plane stays altermagnetic.

\subsection{Model for the Magnetic Normal State}
We next illustrate these statements with a concrete tight-binding model, which we will also use for our explicit calculations of the SDE below. The Hamiltonian is defined on a single breathing kagome lattice with three Mn sublattices, labeled by $\alpha\in\{A,B,C\}$. In momentum space, the normal-state Hamiltonian takes the form,
$
h_{\boldsymbol{k}} = h^{(0)}_{\boldsymbol{k}} + h^{(M)}
$,
where $h^{(0)}_{\boldsymbol{k}}$ describes the kinetic energy, including spin-orbit coupling, and $h^{(M)}$ encodes the magnetic order. 
First, $h^{(0)}_{\boldsymbol{k}}$ acts on the spinor
$
\Psi^{\dagger}_{\boldsymbol k}
=
(
c^\dagger_{\boldsymbol k,A\uparrow},
c^\dagger_{\boldsymbol k,A\downarrow},
c^\dagger_{\boldsymbol k,B\uparrow},
...,
c^\dagger_{\boldsymbol k,C\downarrow}
)
$
and has the block structure\,\cite{PhysRevLett.112.017205,PhysRevB.99.165141,PhysRevResearch.2.033112}
\begin{equation}
h^{(0)}_{\boldsymbol{k}} 
=
\begin{pmatrix}
0 
&
\hat{t}^{\prime\dagger}_{1} + \hat{t}^{\dagger}_{1} e^{-i\boldsymbol{k}\cdot\boldsymbol{a}_{1}} 
&
\hat{t}'_{3} + \hat{t}^{}_{3} e^{i\boldsymbol{k}\cdot\boldsymbol{a}_{3}} 
\\
\hat{t}_{1}^{\prime} + \hat{t}_{1}^{} e^{i\boldsymbol{k}\cdot\boldsymbol{a}_{1}} 
&
0
&
\hat{t}^{\prime\dagger}_{2} + \hat{t}^{\dagger}_{2} e^{-i\boldsymbol{k}\cdot\boldsymbol{a}_{2}} 
\\
\hat{t}^{\prime\dagger}_{3} + \hat{t}^{\dagger}_{3} e^{-i\boldsymbol{k}\cdot\boldsymbol{a}_{3}} 
&
\hat{t}'_{2} + \hat{t}_{2}^{} e^{i\boldsymbol{k}\cdot\boldsymbol{a}_{2}} 
&
0
\end{pmatrix}.
\end{equation}
Here, 
$\boldsymbol{a}_{1} = (-1/2,-\sqrt{3}/2)^T$, 
$\boldsymbol{a}_{2} = (1,0)^T$,
and
$\boldsymbol{a}_{3} = (-1/2,\sqrt{3}/2)^T$
are nearest-neighbor bond vectors, where the lattice constant has been set to unity.

\begin{figure}[!t]
    \centering
    \includegraphics[width=1\linewidth]{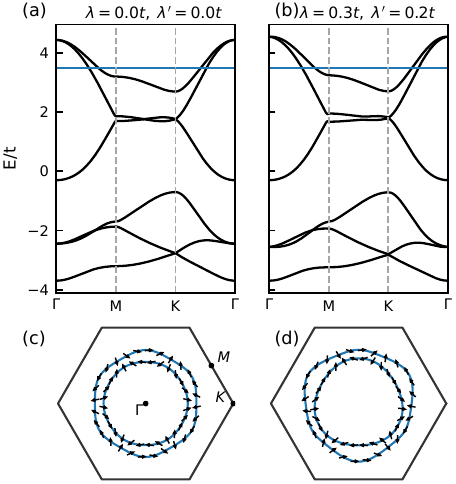}
    \caption{
\textbf{Normal state bands and Fermi surface spin textures for the \textsf{T2} phase.}
(a) Band structure along $\Gamma$–$M$–$K$–$\Gamma$ without spin-orbit coupling ($\lambda = \lambda' = 0)$, showing an altermagnetic spin splitting. (b) Same as (a) but with a staggered spin-orbit coupling on the two kagome triangles $(\lambda\neq\lambda')$. 
(c),(d) Fermi surfaces and Fermi-surface spin textures at the chemical potential indicated in (a) and (b) by a blue line.  
}
    \label{fig:2}
\end{figure}

The matrices $\hat{t}_{\alpha}$ and $\hat{t}'_{\alpha}$ describe nearest-neighbor hopping within the two inequivalent triangles of the breathing kagome lattice and are given by
\begin{equation}
\begin{split}
\hat{t}_{\alpha}
&=
t
- i \lambda\, \boldsymbol{d}_{\alpha}\cdot \boldsymbol{\sigma},
\\
\hat{t}'_{\alpha}
&=
t'
- i \lambda'\, \boldsymbol{d}_{\alpha}\cdot \boldsymbol{\sigma}.
\end{split}
\end{equation}
Here, $\boldsymbol{\sigma}=(\sigma_x,\sigma_y,\sigma_z)$ are the Pauli matrices in spin space, $t$, $t'$ denote spin-independent hopping amplitudes, and $\lambda$, $\lambda'$ parametrize Rashba spin-orbit coupling on the two types of triangles. The bond-orientation vectors entering the spin-orbit term are
$\boldsymbol{d}_{1}= (\sqrt{3}/2,-1/2,0)^T$,
$\boldsymbol{d}_{2}= (0,1,0)^T$, and
$\boldsymbol{d}_{3}= (-\sqrt{3}/2,-1/2,0)^T$ and have been chosen so as to respect the (relativistic) $C_{3v}$ point group. We note that the non-breathing limit corresponds to $t=t'$ and $\lambda=\lambda'$.

Second, the magnetic texture acts on each sublattice as a local exchange field. Its contribution to the Hamiltonian is given by 
\begin{equation}
h^{(M)}
=
\sum_{\alpha}
\big(
\boldsymbol{m}_{\alpha}\cdot \boldsymbol{\sigma}
\big)
\otimes
|\alpha\rangle\langle\alpha|.
\end{equation}
Here, we parametrize the exchange fields as
$\boldsymbol{m}_\alpha
=
m_{0}(\cos\phi_\alpha,\sin\phi_\alpha,0)^T$,
where $m_0$ sets the magnitude of the exchange field and $\phi_\alpha = \varphi_\alpha+\varphi_0$. The angles, $\phi_{\alpha}$, fix the relative orientations of the magnetic moments and are given by $
\varphi_{A}=\pi/2$, $\varphi_{B}=7\pi/6$, and $\varphi_{C}=11\pi/6$. Meanwhile, $\varphi_0$ determines the overall in-plane orientation of the spins, which does not affect the spectrum without spin-orbit coupling; $\varphi_0 = 0$ and $\varphi_0 = \pi/2$ correspond to the \textsf{\textsf{T1}} and \textsf{\textsf{T2}} orders shown in \figref{fig:1}(c) and (d), respectively.

Setting the spin-orbit coupling first to zero, $\lambda = \lambda'=0$, we show the band structure in \figref{fig:2}(a), which clearly reveals a sizable spin splitting (six bands, coming from the three sublattices and two spins) despite the absence of a net magnetization, $\vec{M}=0$, which we have checked by explicit calculation. As can be more clearly seen from the Fermi surfaces in \figref{fig:2}(c), the spin splitting is even in momentum, in line with our symmetry analysis. Furthermore, we display the in-plane components of the spin polarization on the Fermi surfaces, using the \textsf{\textsf{T2}} state as an example. As anticipated, it obeys $[\vec{s}_n(\vec{k})]_{x,y} = [\vec{s}_n(-\vec{k})]_{x,y}$. We refer to the Supplementary Information for an illustration of the spectrum of the \textsf{T1} state and of $[\vec{s}_n(\vec{k})]_z$.

\subsection{Spin-orbit coupling and SDE}
Although Pt is nonmagnetic, it is expected to play a significant role for the strength of spin-orbit coupling in the system.
The key difference in the presence of spin-orbit coupling is that all symmetry operations must be spinful, in the sense that real-space transformations are always accompanied by the associated transformations in spin space. As such, threefold rotational symmetry $C_{3z}^s$ is still present and continues to suppress any in-plane magnetization $M_{x,y}$---irrespective of $\varphi_0$. However, $\hat{s}$ in $[C_{2\hat{s}}||\sigma_v]$ now has to be perpendicular to the mirror plane such that only the \textsf{\textsf{T2}} is invariant under it, while it is broken by the \textsf{\textsf{T1}} state. Consequently, the former is still fully compensated ($\vec{M}=0$) while spin-orbit coupling induces a small out-of-plane magnetization in the latter, $M_z \neq 0$. Such an induced moment is also found in first-principles calculations \cite{PhysRevB.92.144426}. 

In \figref{fig:2}(b) and (d), we show the band structure and Fermi surfaces along with the spin polarizations $[\vec{s}_n(\vec{k})]_{x,y}$, respectively, again for the \textsf{\textsf{T2}} magnetic texture but now with spin-orbit coupling, $\lambda,\lambda'\neq 0$. The main difference is that the product of $\Theta_s$ and $[C_{2z}||E]$ is no longer a symmetry, such that $[\vec{s}_n(\vec{k})]_{x,y}$ and $E_n(\vec{k})$ are not even functions of $\vec{k}$ anymore. 

We emphasize that the breaking of $\Theta_s[C_{2z}||E]$ is essential for the SDE: as the current is odd under this symmetry, its presence would make the critical current along any direction $\hat{\vec{n}}$ and $-\hat{\vec{n}}$ equivalent. 

\section{Proximity effect and Supercurrent}
We next couple this magnetic system to a superconductor via the proximity effect. We describe pairing in the superconductor by a complex order parameter $\Delta(\boldsymbol x)$ and the induced pairing correlations in the altermagnet by $\bar{\Delta}(\boldsymbol x)$. First setting the coupling between the two subsystems to zero, to quartic order, the bare Ginzburg-Landau free energy reads as
\begin{equation}
\begin{split}
\mathcal{F}_0
&=
\sum_{\boldsymbol q}
\Big[
\alpha_{\text{SC}}(\boldsymbol q)\,|\Delta_{\boldsymbol q}|^2
\Big]
+
b_{\text{SC}} \int d^2x\, |\Delta(\boldsymbol x)|^4
\\
&+
\sum_{\boldsymbol q}
\Big[
\alpha_{\text{AM}}(\boldsymbol q)\,|\bar\Delta_{\boldsymbol q}|^2
\Big]
+
b_{\text{AM}} \int d^2x\, |\bar\Delta(\boldsymbol x)|^4
\end{split}
\end{equation}
Here, $\Delta_{\vec{q}}$ ($\bar{\Delta}_{\vec{q}}$) is the Fourier transform of $\Delta(\vec{x})$ ($\bar{\Delta}(\vec{x})$) and the quadratic coefficients are given by
$\alpha_{\text{SC}}(\boldsymbol q)=g^{-1}_{\text{SC}}-\Pi_{\text{SC}}(\boldsymbol q)$ and
$\alpha_{\text{AM}}(\boldsymbol q)=g^{-1}_{\text{AM}}-\Pi_{\text{AM}}(\boldsymbol q)$,
where $g_{\text{SC}}$ and $g_{\text{AM}}$ are effective pairing interactions and
$\Pi_{\text{SC}}(\boldsymbol q)$, $\Pi_{\text{AM}}(\boldsymbol q)$ are the corresponding particle-particle bubbles.

\begin{figure}[!t]
    \centering
    \includegraphics[width=1\linewidth]{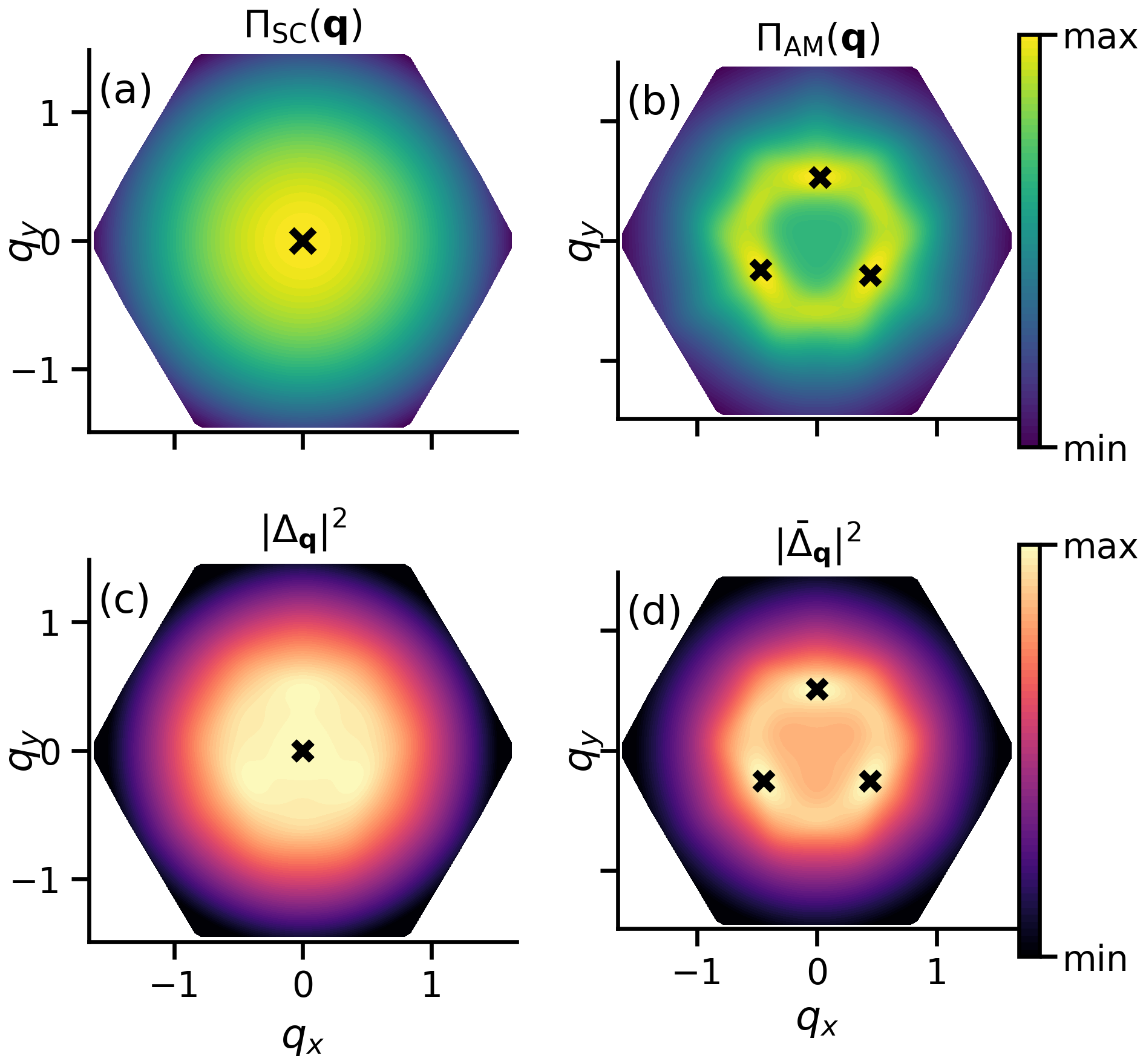}
    \caption{
\textbf{Momentum-dependence of particle-particle bubbles and superconducting order parameters.}
(a) Particle-particle bubble of the superconductor, $\Pi_{\text{SC}}(\boldsymbol q)$, with a single maximum at $\boldsymbol q=0$. 
(b) Particle-particle bubble of the altermagnet, $\Pi_{\text{AM}}(\boldsymbol q)$, showing three $C_3$-related maxima at finite momenta. 
(c) Renormalized order parameter of the superconductor, $|\Delta_{\boldsymbol q}|^{2}$. 
(d) Proximity-induced supercondcuting order parameter in the altermagnet, $|\bar\Delta_{\boldsymbol q}|^{2}$, whose maxima occur at finite momenta, indicating finite-momentum pairing.
    }
    \label{fig:3}
\end{figure}

For a conventional $s$-wave superconductor with time-reversal symmetry in the normal state, $\Pi_{\text{SC}}(\boldsymbol q)$ is even in $\boldsymbol q$ and maximal at $\boldsymbol q=0$ to favor uniform ($\boldsymbol q=0$) pairing in the absence of the altermagnet, see Fig.\,\ref{fig:3}(a). We therefore adopt the approximate form
$\Pi_{\text{SC}}(\boldsymbol q)\approx\Pi_{\text{SC}}(\boldsymbol 0)+\gamma \boldsymbol q^2$
and choose $g_{\text{SC}}$ such that $\alpha_{\text{SC}}(\boldsymbol 0)<0$, as required for superconductivity.

In contrast, the noncollinear altermagnet breaks time-reversal symmetry. Therefore, its particle-particle bubble does not diverge with decreasing temperature, and we can choose the coupling constant $g_{\text{AM}}$ such that the altermagnet is not superconducting on its own, $\alpha_{\text{AM}}(\boldsymbol q)>0$. Another important consequence is that $\Pi_{\text{AM}}(\boldsymbol q)$ is not even in $\vec{q}$ anymore. This can be seen Fig.\,\ref{fig:3}(b), where the maximum is not even located at $\boldsymbol q=0$ anymore but has been shifted to three $\boldsymbol q \neq 0 $ (related by a $C_3$ rotation).
To obtain this result, we computed $\Pi_{\text{AM}}(\boldsymbol q)$ from
\begin{equation}
\Pi_{\text{AM}}(\boldsymbol{q}) = 
T
\sum_{\boldsymbol{k},\omega_{n}}
\text{Tr}
\left[
\hat{\Gamma}
\hat{G}(\boldsymbol{k}_{+}, \omega_{n})
\hat{\Gamma}^{T}
\hat{G}^{T}(-\boldsymbol{k}_{-}, -\omega_{n})
\right],
\end{equation}
where $\boldsymbol k_\pm=\boldsymbol k\pm \boldsymbol q/2$,
$\hat G(\boldsymbol k,\omega)=(i\omega-h_{\boldsymbol k})^{-1}$ is the normal-state Green's function,
$T$ is temperature, and $\omega_n$ are fermionic Matsubara frequencies. For simplicity, we took a spin-singlet, sublattice-diagonal pairing vertex
$\hat{\Gamma}= i\sigma_y\otimes\sum_\alpha|\alpha\rangle\langle\alpha|$.

Next, we couple the two systems via interfacial tunneling, which induces superconducting correlations in the altermagnet. At the level of the free energy, we describe this coupling by
\begin{equation}
\mathcal{F}_c
=
-\int d^2x\,
\Big[
T(\boldsymbol x)\,\bar{\Delta}^*(\boldsymbol x)\Delta(\boldsymbol x)
+\text{c.c.}
\Big],
\end{equation}
where $T(\boldsymbol x)$ denotes the hybridization strength. While moiré effects can give rise to spatial variations, we here retain only the uniform component, $T_0$, since we do not expect any key qualitative changes in our results from additional moiré modulations. The full free energy is then $\mathcal{F}=\mathcal{F}_0+\mathcal{F}_c$.

To obtain analytic expressions for the proximity-induced order parameter, we now assume that both condensates carry a single Cooper-pair momentum $\boldsymbol q_0$, i.e.,
$\Delta_{\boldsymbol q}\propto \delta_{\boldsymbol q,\boldsymbol q_0}$ and
$\bar{\Delta}_{\boldsymbol q}\propto \delta_{\boldsymbol q,\boldsymbol q_0}$.
The free energy then reduces to the effective form, 
\begin{align}
\mathcal{F}_{\text{eff}}(\boldsymbol q)
=&\;
\alpha_{\text{SC}}(\boldsymbol q)\,|\Delta_{\boldsymbol q}|^2
+
\alpha_{\text{AM}}(\boldsymbol q)\,|\bar{\Delta}_{\boldsymbol q}|^2
\\
&\;
-2\,\text{Re}\!\left[T_0\,\Delta^*_{\boldsymbol q}\bar{\Delta}_{\boldsymbol q}\right]
+
b_{\text{SC}}|\Delta_{\boldsymbol q}|^4
+
b_{\text{AM}}|\bar{\Delta}_{\boldsymbol q}|^4 \nonumber.
\end{align}
Minimizing $\mathcal{F}_{\text{eff}}$ yields (to second order in $T_0$) the renormalized superconducting order parameter and the proximity-induced order parameter,
\begin{subequations}
\begin{align}
|\bar{\Delta}^{(0)}_{\boldsymbol{q}}|^{2}
&=
\left(
\frac{
|T_{0}|
}
{
\alpha_{\text{AM}}(\boldsymbol q)
}
\right)^{2}
|\Delta^{(0)}_{\boldsymbol{q}}|^{2} \label{InducedOP}
\\
|\Delta^{(0)}_{\boldsymbol{q}}|^{2}
&=
\text{max}\left\{
0 \ , \
\frac{
1
}
{
2 b_{\text{SC}}
}
\left(
\frac{|T_{0}|^{2}
}
{
\alpha_{\text{AM}}(\boldsymbol q)
}
-
\alpha_{\text{SC}}(\boldsymbol q)
\right)
\right\}. \label{SupercondOP}
\end{align}
\end{subequations}

Three comments about this result are in order: First, we note that the proximity-induced superconducting order parameter is $\propto|T_{0}|^{2}$. Hence, the pairing in the altermagnet vanishes, as expected, if we turn off the proximity coupling, $T_{0}=0$. Second, the order parameter in the superconductor, 
$|\Delta^{(0)}_{\boldsymbol{q}}|^{2}$, is renormalized due to the presence of the altermagnet. Specifically, the renormalized order parameter is determined from the effective quadratic coefficient $\alpha^{\text{eff}}_{\text{SC}}(\boldsymbol q)=\alpha_{\text{SC}}(\boldsymbol q)-(|T_{0}|^{2}/\alpha_{\text{AM}}(\boldsymbol q))$, which as a result of $\alpha_{\text{AM}}$ is not an even function of $\vec{q}$ anymore, see Fig.\,\ref{fig:3}(c). 
Third, this effect is further amplified in the proximity-induced pairing $|\bar{\Delta}^{(0)}_{\boldsymbol{q}}|^{2}$, due to the additional factor of $1/\alpha_{\text{AM}}^2$ in \equref{InducedOP}, which is clearly visible in the form of the more pronounced peaks in \figref{fig:3}(d) compared to \figref{fig:3}(c), which now even occur at non-zero momenta.

\begin{figure}[!t]
    \centering
    \includegraphics[width=1\linewidth]{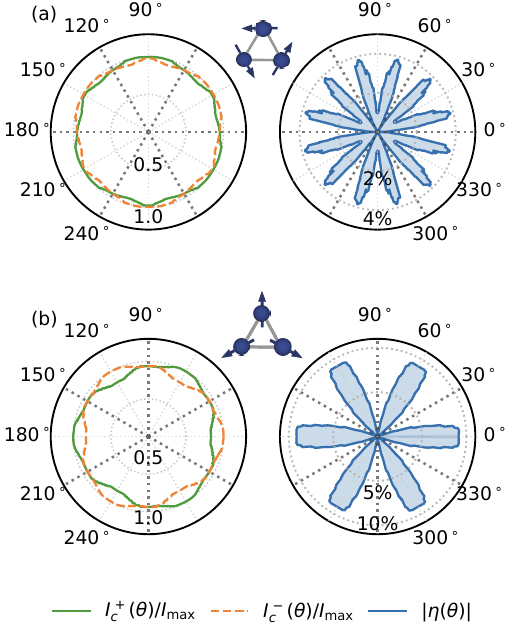}
    \caption{
\textbf{Angular dependence of the superconducting diode effect.}
(a) Angular dependence of the forward and reverse critical currents, $I_c^\pm(\theta)$ (left), and the corresponding diode efficiency, $\eta(\theta)$ (right), for the \textsf{T2} phase. The critical currents are normalized by $I_{\mathrm{max}}=\max_{\theta}\{I_c^\pm(\theta)\}$.
(b) Same quantities for the \textsf{T1} phase. The angular directions for which $\eta(\theta)=0$ differ between \textsf{T1} and \textsf{T2}, providing a possible way to distinguish between the two phases.
    }
    \label{fig:4}
\end{figure}

Finally, we can also obtain an expression for the supercurrent as the gradient of our effective free energy, 
\begin{equation}
\begin{split}
 \boldsymbol J(\boldsymbol q)
&=
2e\,\partial_{\boldsymbol q}
\mathcal{F}_{\text{eff}}(\boldsymbol q)
\\
&=
-2e
\left[
(\partial_{\boldsymbol q}\Pi_{\text{SC}})\,|\Delta^{(0)}_{\boldsymbol{q}}|^2
+
(\partial_{\boldsymbol q}\Pi_{\text{AM}})\,|\bar{\Delta}^{(0)}_{\boldsymbol{q}}|^2
\right].
\label{Eq9}
\end{split}    
\end{equation}
We see that the supercurrent has, as expected, two contributions. The first contribution is the supercurrent carried by the superconductor, while the second contribution is carried by the proximitized altermagnet. 
Although finite-momentum pairing is not our main focus here, we note that the equilibrium value $\vec{q}=\vec{q}_0$ of the system is not necessarily equal to the extremum (one of the degenerate extrema) of $\Pi_{\text{SC}}$ and $\Pi_{\text{AM}}$ (which might even be different) but instead obeys $\vec{J}(\vec{q}_0)=0$.
In the following, we will discuss in detail how this particular form of the supercurrent in \equref{Eq9} can lead to a superconducting diode effect. 

\section{Supercurrent diode effect}
\subsection{General considerations}
We are now in the position to analyze the SDE, i.e., the emergence of different critical currents in the forward and reverse current bias direction. As a starting point, it is instructive to discuss its microscopic origin through the two current contributions in Eq.~\eqref{Eq9}.

The first contribution, associated with the superconductor, is given by 
$
\boldsymbol J_{\text{SC}}(\boldsymbol q)
=
-2e(\partial_{\boldsymbol q}\Pi_{\text{SC}})
\,
|\Delta^{(0)}_{\boldsymbol{q}}|^2
$
. 
Here, the first factor is given by 
$
\partial_{\boldsymbol q}\Pi_{\text{SC}}
=
2\gamma\boldsymbol{q}
$
and is therefore strictly odd in $\boldsymbol q$. Without the renormalization $\propto |T_0|^2$ in \equref{SupercondOP}, $|\Delta^{(0)}_{\boldsymbol{q}}|$ is even in $\vec{q}$ and we have $\boldsymbol J_{\text{SC}}(\boldsymbol q) =  -\boldsymbol J_{\text{SC}}(-\boldsymbol q)$ such that the current is reciprocal. The aforementioned proximity-induced renormalization, however, introduces an odd contribution to $|\Delta^{(0)}_{\boldsymbol{q}}|$ such that the maximal $|\vec{J}_{\text{SC}}(\boldsymbol q)|$ with $\vec{J}_{\text{SC}}(\boldsymbol q)$ pointing along some direction $\hat{\vec{n}}$ and along $-\hat{\vec{n}}$ are in general not identical. 

The non-reciprocal fraction of the current is expected to be larger in the second, altermagnetic, contribution to the current, $\boldsymbol J_{\text{AM}}(\boldsymbol q)=-2e
(\partial_{\boldsymbol q}\Pi_{\text{AM}})\,|\bar{\Delta}^{(0)}_{\boldsymbol{q}}|^2$.
In this case, the key difference already appears at the level of the particle-particle bubble. Namely, 
$\Pi_{\text{AM}}(\boldsymbol{q})\neq\Pi_{\text{AM}}(-\boldsymbol{q})$ and,
as a result, also $\partial_{\boldsymbol q}\Pi_{\text{AM}}$ acquires a contribution that is even in $\boldsymbol{q}$, which already induces a SDE. On top of this, the antisymmetric part in the proximity-induced superconducting order parameter $|\bar{\Delta}^{(0)}_{\boldsymbol{q}}|^2$ is further enhanced compared to $|\Delta^{(0)}_{\boldsymbol{q}}|^2$, as already mentioned above.
The current-momentum relation is therefore no longer antisymmetric, $\boldsymbol J_{\text{AM}}(\boldsymbol q)\neq -\boldsymbol J_{\text{AM}}(-\boldsymbol q)$, and we expect that the altermagnetic contribution can result in a sizable SDE.

\subsection{Angular dependence}
To quantify our considerations, we will now numerically evaluate the diode efficiency and its angular dependence for our setup. We first define the critical currents in the forward and reverse directions,
$I_c^+(\theta)
=
\max_{\mathbf q}
\{
|\boldsymbol{J}(\boldsymbol{q})| \text{ with } \boldsymbol{J}(\boldsymbol{q}) \parallel \hat{\boldsymbol{n}}(\theta)
\}$
and
$
I_c^-(\theta) = I_c^+(\theta +\pi) 
$ 
where $\hat{\mathbf n}(\theta)=(\cos\theta,\sin\theta)$ denotes the in-plane direction of the applied current bias.
The diode efficiency is then given by 
$
\eta(\theta)
=
\frac{|I_c^+(\theta)-I_c^-(\theta)|}
{I_c^+(\theta)+I_c^-(\theta)}
$.

Our results for $\eta(\theta)$ are shown in Fig.\,\ref{fig:4}(a) and (b) for the \textsf{T1} and \textsf{T2} phase, respectively. For the parameters used in our simulations (see Supplemental Information for details), we find that the \textsf{T2} phase reaches a maximal diode efficiency of $\sim 4\%$, whereas the \textsf{T1} phase reaches a larger value of $\sim 9\%$. We highlight the nonzero SDE in the \textsf{T2} phase despite zero net magnetization---a distinctive feature of this form of magnetic order. As already mentioned in the introduction, this differentiates our setup from many other superconducting diodes \cite{Lin2022,FerromagneticProximity,le2024superconducting,wu2022fieldfree}, with finite net magnetization.

In addition to the maximum $\eta$, an interesting prediction of our simulations is that the diode efficiency exhibits a characteristic multi-lobe pattern that provides a way to distinguish between the \textsf{T1} and \textsf{T2} phases experimentally using angle-dependent critical-current measurements \cite{SunbeamSetup}. 
In particular, in the \textsf{T2} phase, we find that the diode efficiency vanishes for
$\theta\in\{0^{\circ},60^{\circ},120^{\circ},180^{\circ},240^{\circ},300^{\circ}\}$. In contrast, in the \textsf{T1} phase, the diode efficiency vanishes for
$\theta\in\{30^{\circ},90^{\circ},150^{\circ},210^{\circ},270^{\circ},330^{\circ}\}$. This naturally follows from the symmetries since $\eta$ is required to vanish for directions $\hat{\vec{n}}$ perpendicular to mirror planes or if it lies in a magnetic mirror plane \cite{banerjee2024altermagnetic}. Consequently, the $\sigma_v^s \Theta_s$ ($\sigma_v^s$) symmetry of the \textsf{T1} (\textsf{T2}) phase leads to $\eta(\theta) = 0$ for $\theta = 90^\circ$ ($\theta = 0^\circ$) and symmetry-related directions.

We finally note that we assumed in the above calculation that the superconductor in its current-carrying state always reaches the most stable solution. 
However, if there are three $C_{3z}$-related non-zero equilibrium momenta, it is possible that a vestigial order parameter, associated with choosing one of the three $\vec{q}_0$, survives beyond the critical current; the superconductor then only ``explores'' the stable vicinity of $\vec{q}_0$ \cite{scammell_theory_2022,b7rh-v7nq,2025arXiv251122796B}, leading to a $C_{3z}$-breaking $\eta(\theta)$.

\begin{figure}[!t]
    \centering
    \includegraphics[width=1\linewidth]{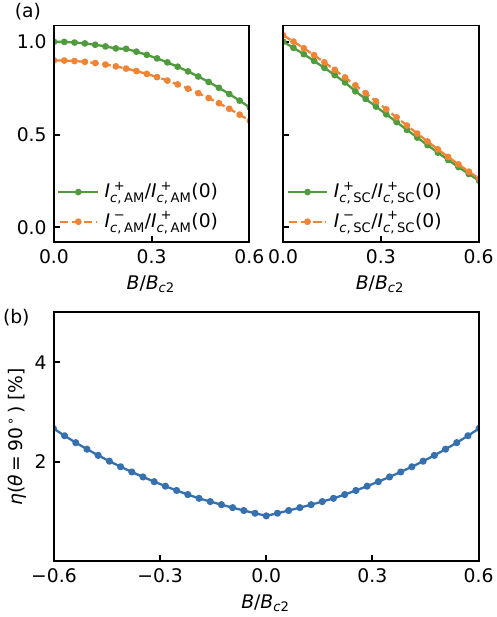}
    \caption{
\textbf{Magnetic-field dependence of the superconducting diode effect.}
(a) Critical currents as a function of out-of-plane magnetic field, $B$, for the altermagnetic [left, highly nonreciprocal, from second term in \equref{Eq9}] and the superconducting [right, nearly reciprocal, first term in \equref{Eq9}] contributions. The critical currents are normalized by $I_{c,\mathrm{AM}}^{+}(B=0)$ and $I_{c,\mathrm{SC}}^{+}(B=0)$, respectively. The current bias is applied along $\theta=90^\circ$ and the critical field is denoted by $B_{c2}$.
(b) Diode efficiency, $\eta$, as a function of the magnetic field, $B$. The efficiency increases with $B$ as the reciprocal supercurrent contribution is progressively suppressed, which enhances the relative weight of the nonreciprocal supercurrent carried by the altermagnet.   
    }
    \label{fig:5}
\end{figure}

\section{Discussion}
In summary, we provided a detailed symmetry analysis of the \textsf{T1} and \textsf{T2} magnetic states of the (111) Mn$_3$Pt heterostructure of \cite{PartnerExperimentalPaper}, with and without spin-orbit coupling. We further performed explicit tight-binding model calculations on the kagome lattice, which allowed us to illustrate the characteristic altermagnetic spin splitting of the electronic bands. Building on this description, we identified a possible mechanism for the superconducting diode effect that emerges when Mn$_3$Pt is proximitized by a conventional $s$-wave superconductor and computed the directional dependence of the diode efficiency, $\eta(\hat{\vec{n}})$. While both magnetic orders yield a diode effect, only \textsf{T2} remains fully compensated, i.e., there is a symmetry that ensures vanishing net magnetization. 

One interesting tuning parameter in the experiment \cite{PartnerExperimentalPaper}, which we have not discussed yet, is the external magnetic field, $B$, applied perpendicular to the kagome plane. 
We first note that a linear coupling between $B$ and the \textsf{T2} magnetic order parameter is prohibited by the $\sigma_v^s$ symmetry; furthermore, it also has to vanish without spin-orbit coupling for the \textsf{T1} configuration, as a result of $\Theta_s$ combined with $[C_{2z}||E]$, and is thus likely small. As such, we will assume that $B$ does not significantly affect the magnetic order and that the diode effect's $B$-dependence primarily comes from the  suppression of pairing with the orbital effect of $B$.  
This is also consistent with experiment, where the applied magnetic field does not flip the polarity of the magnetic order, no hysteresis is observed, and the diode efficiency was observed to be approximately even in $B$.

Furthermore, $\eta$ is found to increase at small fields.
To explain this behavior theoretically, we will assume that $B$ suppresses the nearly reciprocal contribution to the critical current in the superconductor more efficiently than the significantly non-reciprocal contribution in the proximitized altermagnet. In particular, for the velocity contribution [$\partial_{\vec{q}}\Pi_{\text{SC,AM}}$ in \equref{Eq9}], this is not unnatural since time-reversal is already broken in the normal state associated with $\Pi_{\text{AM}}$, such that spatial inhomogeneities (here due to a spatially varying order parameter, e.g., vortices) can even enhance pairing \cite{PhysRevB.108.054510,PhysRevB.111.L100502,2025arXiv251019943S}. 
As long as the suppression of $\vec{J}_{\text{SC}}$ is stronger than for $\vec{J}_{\text{AM}}$, increasing $B$ leads to an increase of the \textit{relative} weight of the non-reciprocal channel and, hence, 
to an enhanced diode efficiency. We demonstrate this in \figref{fig:5} using a phenomenological model (see Methods for details), but leave a systematic study of the impact of the orbital field as future work. 

We remark that our calculated diode efficiencies for the \textsf{T2} phase in \figref{fig:4}(a) are in rough agreement with experimentally reported values \cite{PartnerExperimentalPaper}. For the \textsf{T1} phase, our calculations show an enhancement of the diode efficiency compared to the \textsf{T2} phase, which is again qualitatively consistent with the experiment. However, we caution that the absolute values of diode efficiencies are generally highly parameter sensitive; furthermore, there are likely additional contributions to the diode effect arising from vortex physics, particularly at elevated temperatures $T$ and in finite magnetic fields $B$. We believe that this is the main missing ingredient needed to explain the high efficiencies observed experimentally in the \textsf{T2} state at larger $T$ and $B$.

Apart from the quantitative differences in the maximal diode efficiencies, a key qualitative difference in the directional dependencies of $\eta$ between the \textsf{T1} and \textsf{T2} phases are the directions along which $\eta(\hat{\vec{n}})$ vanishes, see dashed lines in \figref{fig:4}(a) and (b). We emphasize that this behavior is expected to be rather robust and not specific to our modeling, since it follows from symmetries. It demonstrates how future directional-dependent transport studies would provide additional non-trivial checks of our proposed theory and could further be used to differentiate between the two competing magnetic orders.  
In fact, Mn$_3$Pt could in principle also host other magnetic orders, which might become favorable in the heterostruture. These would then also be visible in angle-resolved transport. Most notably, the competing \textsf{T3} phase \footnotemark[\value{footnote}] is related to the \textsf{T1} order by a $30^\circ$ rotation (along the $z$ direction) and, thus, descends from the same parent altermagnet. In the presence of spin-orbit coupling, it will also lead to a diode effect, however, like the \textsf{T1} state, with small but finite out-of-plane magnetization. As it breaks both $\sigma_v^s$ and $\Theta\sigma_v^s$, $\eta(\hat{\vec{n}})$ will now vanish along generic directions.

More broadly speaking, we hope that our work will lay the theoretical foundation for future realizations of the altermagnetic superconducting diode effect in proximity-coupled magnet-superconductor hybrid systems, complement alternative approaches to superconducting diodes 
\cite{Lin2022,FerromagneticProximity,le2024superconducting,wu2022fieldfree,scammell_theory_2022,zgnk-rw1p,chen2025finitemomentumsuperconductivitychiralbands,yoon2025quartermetalsuperconductivity,ingla2025,telkamp2025voltagetunablefieldfreejosephsondiode,PhysRevB.110.104510} without external magnetic fields, and enable novel applications in hybrid quantum devices \cite{PhysRevApplied.21.064029,dirnegger2025nonreciprocalquantuminformationprocessing,yu2025enhancedsuperconductingdiodeeffect,pitavidal2025novelqubitshybridsemiconductorsuperconductor}.

\section*{Methods}
To describe the impact of the magnetic field on the SDE, we will adopt a phenomenological approach. Concretely, for each ``channel", $j\in\{\text{SC},\text{AM}\}$ we introduce
two suppression factors: First, $f_{j}(B)$, which controls the pairing amplitude, with $f_{j}(0)=1$ and
$f_{j}(B_{c2})=0$, where $B_{c2}$ is the upper critical field of the superconductor. 
Second, $g_{j}(B)$, which controls the ability of the channel to carry a supercurrent,
with $g_{j}(0)=1$ and $g_{j}(B_{\text{tr}})=0$, where $B_{\text{tr}}$ is the
magnetic field at which transport becomes resistive. 
These phenomenological factors rescale the pairing amplitudes, 
$
\Delta^2(B)=f_{\text{SC}}(B)\Delta^2(0)
$
and
$
\bar\Delta^2(B)=f_{\text{AM}}(B)\bar\Delta^2(0)
$, 
as well as the current response via, 
$
\alpha_{\text{SC}}(B)=f_{\text{SC}}(B)\alpha_{\text{SC}}(0)
$
and
$
\alpha_{\text{AM}}(B)=f_{\text{AM}}(B)\alpha_{\text{AM}}(0)
$.
The two contributions to the supercurrent  then inherit the parametric rescalings, 
$
J_{\text{SC}}(B)\propto f_{\text{SC}}(B) g_{\text{SC}}(B)
$
and
$
J_{\text{AM}}(B)\propto f_{\text{AM}}(B)\,g_{\text{AM}}(B)
$.
Thus the relative importance of the non-reciprocal altermagnetic contribution is governed by the
ratio $[f_{\text{AM}}(B)g_{\text{AM}}(B)]/[f_{\text{SC}}(B)g_{\text{SC}}(B)]$.
In particular, if the magnetic field suppresses the supercurrent in the superconductor more
strongly than the supercurrent in the proximitized altermagnet, the non-reciprocal
altermagnetic contribution will grow in relative importance. The result is an increase of the 
diode efficiency with magnetic field. 

For the magnetic field dependence of the order parameters, we took the phenomenological form
$
f_{\text{AM/SC}}(B)
=
[
1-(|B|/B_{c2})^{p_{\text{AM/SC}}}
]^{\alpha_{\text{AM/SC}}}
$ 
for the suppression of the superconducting order parameters, 
and
$g_{\text{AM/SC}}(B)
=
[
1-(|B|/B_{\mathrm{tr}})^{p'_{{\text{AM/SC}}}}
]^{\alpha'_{{\text{AM/SC}}}}$
for the suppression of the stiffness contribution.
For the results in \figref{fig:5}, we took 
$B_{c2}=1$
and 
$B_{\text{tr}}=0.9$, which accounts for the assumption
that transport can become resistive before the superconducting order parameters are destroyed. 
We further set, 
$(p_{\text{SC}},\alpha_{\text{SC}})=(p_{\text{AM}},\alpha_{\text{AM}})=(2,1)$,
so that both superconducting order parameters are suppressed in a comparable way. 
Lastly, we set 
$(p'_{\text{SC}},\alpha'_{\text{SC}})=(1,1)$ and
$\alpha'_{\text{AM}}=0$, 
which accounts for a stronger suppression of the supercurrent contribution in the superconductor 
compared to the altermagnet.

\begin{acknowledgments}
M.S.S., C.S., and S.M. acknowledge important discussions with J.~Shabani. 
M.S.S.~thanks S.~Banerjee for discussions and previous collaborations on the diode effect and altermagnetism. He further acknowledges discussions with R.~Fernandes and J.A.~Sobral. C.S. acknowledges support from the Louisiana Board of Regents. 
S.M. is supported by ANRF (formerly SERB) Core Research Grant (CRG/2023/008193), Government of India.
\end{acknowledgments}

\onecolumngrid

\newpage

\clearpage
\setcounter{page}{1}
\renewcommand{\thepage}{\arabic{page}}

\setcounter{section}{0}
\renewcommand{\thesection}{S\arabic{section}}
\renewcommand{\thesubsection}{S\arabic{section}.\arabic{subsection}}
\renewcommand{\thesubsubsection}{S\arabic{section}.\arabic{subsection}.\arabic{subsubsection}}

\setcounter{figure}{0}
\setcounter{table}{0}
\renewcommand{\thefigure}{S\arabic{figure}}
\renewcommand{\thetable}{S\arabic{table}}

\setcounter{equation}{0}
\renewcommand{\theequation}{S\arabic{equation}}

\begin{center}
\large{\bf Supplemental Information \\}
\end{center}
\begin{center}Constantin Schrade$^{1}$, Sujit Manna$^{2}$, and Mathias S. Scheurer$^{3}$
\\
{\it $^{1}$Hearne Institute of Theoretical Physics, Department of Physics \& Astronomy, Louisiana State University, Baton Rouge LA 70803, USA}\\
{\it $^{2}$Department of Physics, Indian Institute of Technology Delhi, Hauz Khas, New Delhi 110016, India}
\\
{\it $^{3}$Institute for Theoretical Physics III, University of Stuttgart, 70550 Stuttgart, Germany}
\end{center}

\section{Simulation parameters}
In this section, we provide the system parameters for the simulation results presented in the main text. 
The parameters for Fig.~2 and Fig.~3 of the main text are given in Table~\ref{table1} and \ref{table2}, respectively. The parameters used for Fig.~4 of the main text
are the same as for Fig.~3. 

\begin{table}[h!]
\centering
\renewcommand{\arraystretch}{1.1}
\begin{tabular}{@{}l|cccccc@{}}
\toprule
Panel & $t'$ & $\lambda$ & $\lambda'$ & $m_0$ & $\mu$ & $(\varphi_A,\varphi_B,\varphi_C)$ \\
\midrule
(a),(c) & $t$ & $0$ & $0$ & $1.7t$ & $3.5t$ & $(\pi,5\pi/3,7\pi/3)$ \\
(b),(d) & $t$ & $0.3t$ & $0.2t$ & $1.7t$ & $3.5t$ & $(\pi,5\pi/3,7\pi/3)$ \\
\bottomrule
\end{tabular}
\caption{\textbf{Simulation parameters for Fig.~2 of the main text.}}
\label{table1}
\end{table}
\vspace{-0.3cm}
\begin{table}[h!]
\centering
\renewcommand{\arraystretch}{1.1}
\begin{tabular}{@{}lcccccccccccc@{}}
\toprule
 & $t'$ & $\lambda$ & $\lambda'$ & $m_0$ & $\mu_{\text{AM}}$ & $T$ &
 $\Pi_{\rm SC}(\boldsymbol{0})$ & $\gamma$ & $T_0$ & $g_{\text{SC}}^{-1}$ & $g_{\rm AM}^{-1}$ & $(\varphi_A,\varphi_B,\varphi_C)$ \\
\midrule
&  $t$ & $0.3t$ & $0.2t$ & $1.7t$ & $3.5t$ & $0.02t$
& $0.49t^{-1}$ & $-0.16t^{-1}$ & $0.9t^{-1}$ & $0.44t^{-1}$ & $3.6422t^{-1}$ & $(\pi,5\pi/3,7\pi/3)$ \\
\bottomrule
\end{tabular}
\caption{\textbf{Simulation parameters for Fig.~3 of the main text.} 
For this parameter set, the superconductor satisfies
$\alpha_{\text{SC}}(\boldsymbol{0})=g_{\text{SC}}^{-1}-\Pi_{\text{SC}}(0)=-0.05t^{-1}<0$. 
Moreover, the altermagnet is nonsuperconducting in the absence of the proximity effect, 
$\min_{\boldsymbol{q}}\{\alpha_{\text{AM}}(\boldsymbol{q})\}\approx1.95t^{-1}>0$.
Moreover, the proximity coupling is sufficiently weak so that $|T_0|/\alpha_{\rm AM}(\boldsymbol{q})\le r_{\max}\equiv 0.5$ is satisfied for all $\boldsymbol{q}$, because
$\max_{\boldsymbol{q}}\{|T_0|/\alpha_{\rm AM}(\boldsymbol{q})\}\approx0.46$.
}
\label{table2}
\end{table}

\section{Additional details on the normal state spectrum and the spin texture}
\begin{figure}[!h]
    \centering
    \includegraphics[width=0.8\linewidth]{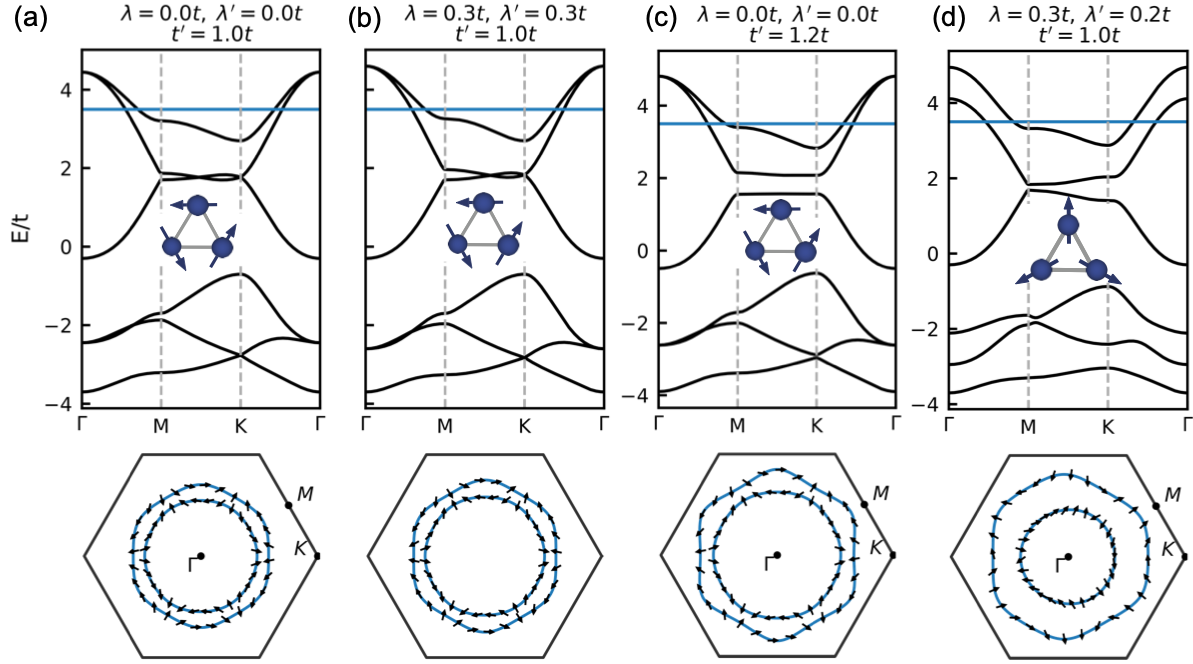}
    \caption{
\textbf{Additional data on the normal state bands and in-plane Fermi surface spin textures of Mn$_3$Pt.}
Top row: band structure along $\Gamma$-$M$-$K$-$\Gamma$ for the \textsf{T2} phase (a,b,c) and the \textsf{T1} phase (d).
Bottom row: corresponding Fermi surface spin textures at the chemical potential indicated by the blue horizontal line.
(a) \textsf{T2} phase without spin-orbit coupling, $\lambda=\lambda'=0$ (identical to Fig.~2(a) in the main text).
(b) \textsf{T2} phase with finite and equal spin-orbit coupling, $\lambda=\lambda'=0.3t$. 
(c) \textsf{T2} phase without spin-orbit coupling, $\lambda=\lambda'=0$, and unequal spin-independent hoppings, $t'=1.2t$. 
(d) \textsf{T1} phase with finite and unequal spin-orbit coupling, $\lambda=0.3t$, $\lambda'=0.2t$. 
    }
    \label{fig:1SI}
\end{figure}

\begin{figure}[!h]
    \centering
    \includegraphics[width=0.9\linewidth]{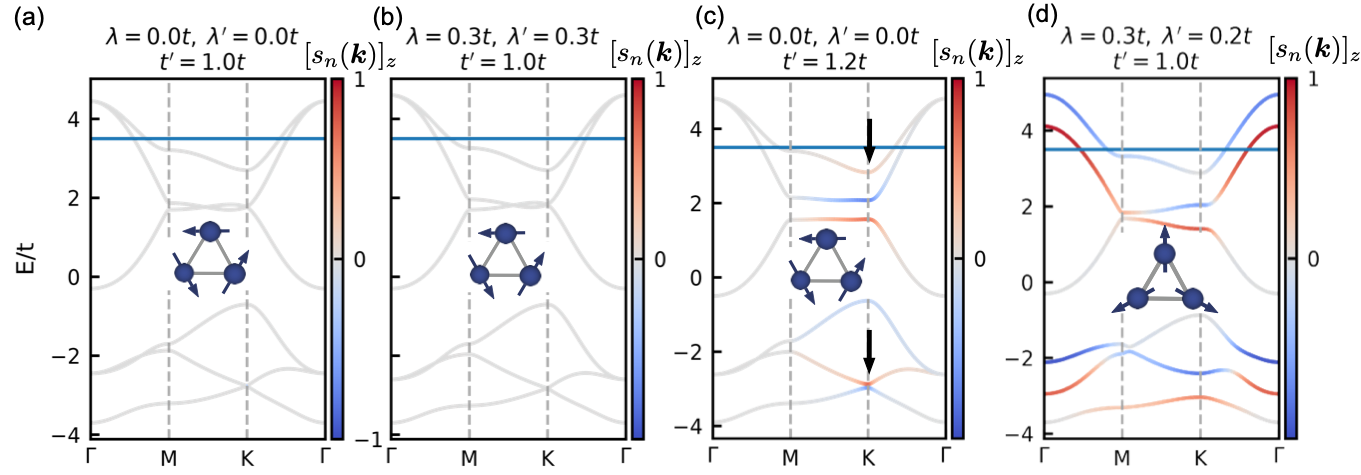}
    \caption{
\textbf{Normal state bands of Mn$_3$Pt with out-of-plane spin polarization.}
Band structure along $\Gamma$-$M$-$K$-$\Gamma$ for the \textsf{T2} phase (a,b,c) and the \textsf{T1} phase (d), with the out-of-plane spin expectation value $[s_n(\boldsymbol{k})]_z$ shown on the color scale.
System parameter are identical to those used in the panels of Fig.~\ref{fig:1SI}.
(a) \textsf{T2} phase without spin-orbit coupling, $\lambda=\lambda'=0$. 
(b) \textsf{T2} phase with finite and equal spin-orbit coupling, $\lambda=\lambda'=0.3t$.
(c) \textsf{T2} phase without spin-orbit coupling, $\lambda=\lambda'=0$, and unequal spin-independent hoppings, $t'=1.2t$. Black arrows highlight the emergence of a finite out-of-plane spin polarization near the corners of the Brillouin zone.
(d) \textsf{T1} phase with finite and unequal spin-orbit coupling, $\lambda=0.3t$, $\lambda'=0.2t$.
    }
    \label{fig:2SI}
\end{figure}

\begin{figure}[!h]
    \centering
    \includegraphics[width=0.85\linewidth]{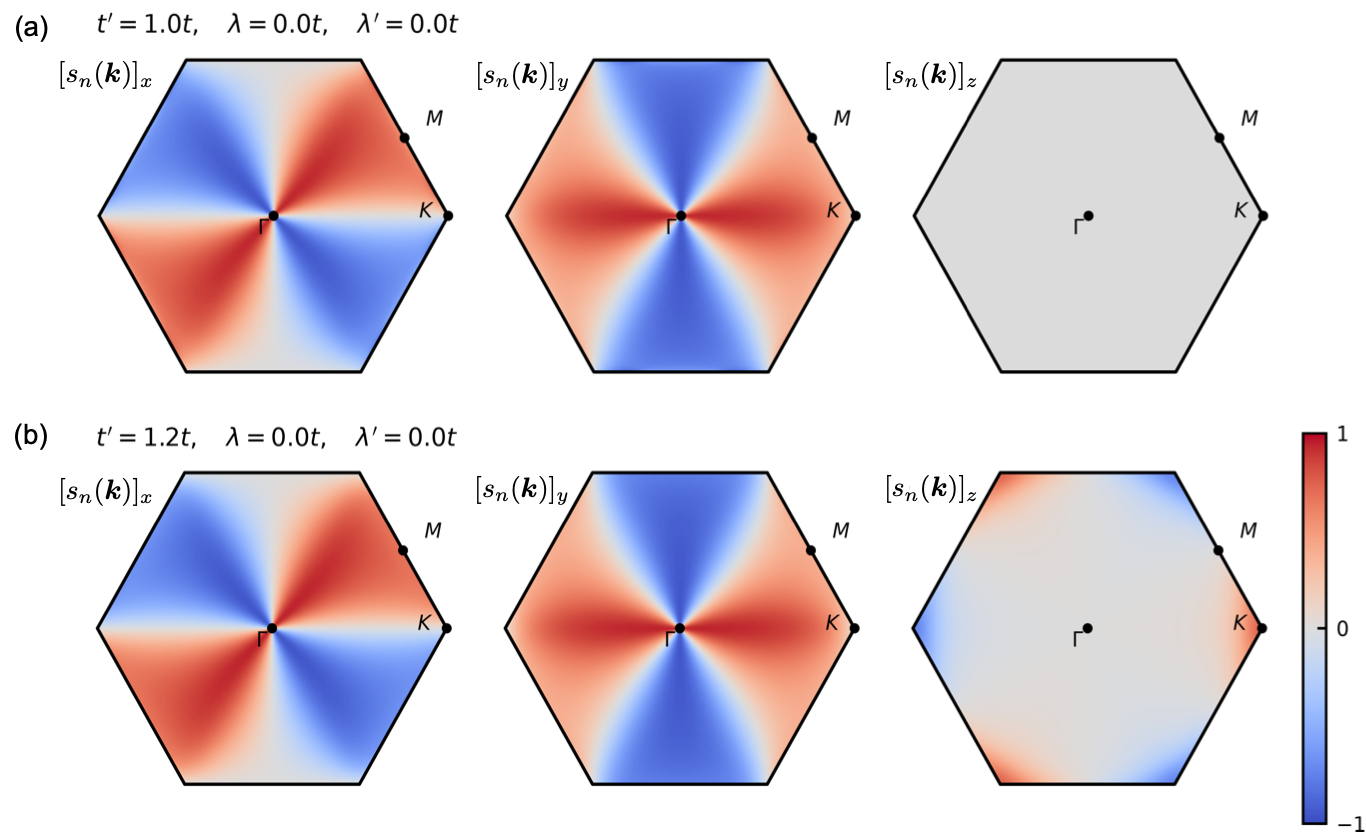}
\caption{
\textbf{Spin expectation values in the \textsf{T2} phase of Mn$_3$Pt over the full Brillouin zone.}
(a) Non-breathing limit,  $t'=1.0t$, with vanishing spin-orbit coupling, $\lambda=\lambda'=0$. Only the in-plane components, $[s_n(\boldsymbol{k})]_{x,y}$, are finite and are even under $\boldsymbol{k}\to-\boldsymbol{k}$. The out-of-plane component, $[s_n(\boldsymbol{k})]_z$, vanishes.
(b) Breathing case, $t'=1.2t$, with  vanishing spin-orbit coupling, $\lambda=\lambda'=0$. A finite out-of-plane spin polarization, $[s_n(\boldsymbol{k})]_z$, emerges, which is odd under $\boldsymbol{k}\to-\boldsymbol{k}$ and thus antialtermagnetic.
High-symmetry points, $\Gamma$, $M$, and $K$, are marked in the Brillouin zone.}
    \label{fig:3SI}
\end{figure}

\begin{figure}[!h]
    \centering
    \includegraphics[width=0.9\linewidth]{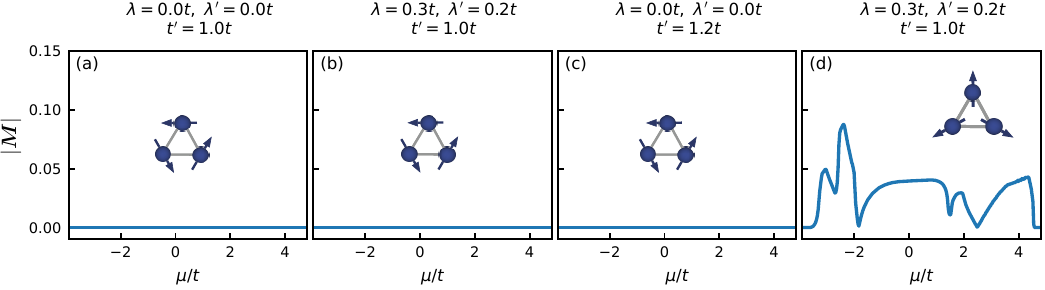}
\caption{
\textbf{Magnitude of the magnetization of Mn$_3$Pt as a function of the chemical potential} 
(a) In the  \textsf{T2} phase without spin-orbit coupling, $\lambda=\lambda'=0$. 
(b) In the \textsf{T2} phase with finite and unequal spin-orbit coupling, $\lambda=0.3t$ and $\lambda'=0.2t$.
(c) In the \textsf{T2} phase without spin-orbit coupling, $\lambda=\lambda'=0$, and unequal spin-independent hoppings, $t'=1.2t$. 
(d) In the \textsf{T1} phase with finite and unequal spin-orbit coupling, $\lambda=0.3t$, $\lambda'=0.2t$.}
    \label{fig:4SI}
\end{figure}

In this section, we provide additional details on the normal-state spectrum and the magnetic properties of Mn$_3$Pt. Specifically, Fig.\,\ref{fig:1SI} shows additional data on the normal-state spectrum, Figs.\,\ref{fig:2SI} and \ref{fig:3SI} display additional results on the momentum-space spin textures, and Fig.\,\ref{fig:4SI} presents the total magnetization.
\\
\\
First, we discuss Fig.\,\ref{fig:1SI}. 
Here, we note that when the spin-orbit couplings on the two kagome triangles are equal, $\lambda=\lambda'$, the resulting Fermi surface is symmetric under momentum reversal, as shown in Fig.\,\ref{fig:1SI}(b). 
The same symmetry persists when the spin-independent hopping amplitudes are different, $t\neq t'$, provided that spin-orbit coupling is absent, $\lambda=\lambda'=0$, as shown in Fig.\,\ref{fig:1SI}(c). 
In both cases, the equivalence of forward and reverse momentum directions precludes a superconducting diode effect. By contrast, in the $\textsf{T1}$ phase with unequal spin–orbit couplings, $\lambda\neq\lambda'$, the Fermi surface generically becomes asymmetric under momentum reversal, as shown in Fig.\,\ref{fig:1SI}(d). This inequivalence between forward and backward directions is the microscopic origin of the superconducting diode effect.
\\
\\
Second, we discuss Fig.\,\ref{fig:2SI}. 
Here, we show the out-of-plane component of the momentum-space spin texture, $[s_n(\boldsymbol{k})]_z$, plotted on a color scale for the normal-state spectra of Fig.\,\ref{fig:1SI}.
For zero spin-orbit coupling, see Fig.\,\ref{fig:2SI}(a), or when it is equal on the two kagome triangles, $\lambda=\lambda'$, see Fig.\,\ref{fig:2SI}(b), the out-of-plane spin component vanishes, $[s_n(\boldsymbol{k})]_z=0$, as long as $t=t'$. In contrast, when the two kagome triangles are inequivalent by unequal spin-independent hoppings, $t\neq t'$, even in the absence of spin–orbit coupling, the system develops a finite out-of-plane antialtermagnetic component, as shown in Fig.\,\ref{fig:2SI}(c).
Lastly, when the inequivalence between the kagome triangles arises from unequal spin–orbit couplings, a finite, $\vec{k}$-local out-of-plane spin expectation value emerges in the \textsf{T1} phase, 
as shown in Fig.\,\ref{fig:2SI}(d).
\\
\\
Third, we discuss Fig.\,\ref{fig:3SI}. 
Here, we show the in-plane components, $[s_n(\boldsymbol{k})]_{x,y}$, and the out-of-plane component, $[s_n(\boldsymbol{k})]_z$, of the spin texture over the full Brillouin zone. 
When the two kagome triangles are inequivalent by unequal spin-independent hoppings, $t\neq t'$, in the absence of spin–orbit coupling, $\lambda\neq\lambda'$, a finite out-of-plane spin component can develop.
This component is antialtermagnetic, because it is odd under momentum reversal, $\boldsymbol{k}\to -\boldsymbol{k}$.
\\
\\
Fourth, we discuss Fig.\,\ref{fig:4SI}. Here, we show the magnitude of the total magnetization as a function of the chemical potential. We note that the total magnetization vanishes in the \textsf{T2} phase for the parameters shown in Fig.\,\ref{fig:4SI}(a-c). In comparison, it can be finite in the \textsf{T1} phase, as shown in Fig.\,\ref{fig:4SI}(d).
\\
\\

\vspace*{3cm}

\newpage

\section{Computation of the critical current}
In this section, we provide details on the description of superconductivity and the computation of the critical current in a  
heterostructure comprised of a superconductor (SC) and altermagnet (AM). 
\\
\\
We start from a Hamiltonian of the form $\mathcal{H} = \mathcal{H}_{\text{SC}} + \mathcal{H}_{\text{AM}} + \mathcal{H}_c$. Here, $\mathcal{H}_{\text{SC}}$ describes the superconductor, which we model on the mean-field level as
\begin{equation}
    \mathcal{H}_{\text{SC}} = \sum_{\boldsymbol{k},\sigma} \xi_{\boldsymbol{k}} c^\dagger_{\boldsymbol{k},\sigma} c^{}_{\boldsymbol{k},\sigma} + \sum_{\boldsymbol{k},\boldsymbol{q}} \left[ \Delta_{\boldsymbol{q}} c^\dagger_{\boldsymbol{k} +\boldsymbol{q}/2} is_y c^\dagger_{-\boldsymbol{k} +\boldsymbol{q}/2} + \text{H.c.} \right] + \sum_{\boldsymbol{q}} \frac{|\Delta_{\boldsymbol{q}}|^2}{g_{\text{sc}}}.
\end{equation}
Here, $\xi_{\boldsymbol{k}} = \xi_{-\boldsymbol{k}}$ is the normal-state dispersion of the active band hosting superconductivity, $c^\dagger_{\boldsymbol{k},\sigma}$ are the associated electronic band creation operators, and $g_{\text{sc}}$ is the coupling constant.
Integrating out the fermions and expanding in powers of $\Delta_{\boldsymbol{q}}$, we obtain the free-energy expression,
\begin{equation}
    \mathcal{F}_{\text{SC}} \sim \sum_{\boldsymbol{q}} \left[ \frac{1}{g_{\text{sc}}} - \Pi_{\text{SC}}(\boldsymbol{q}) \right] |\Delta_{\boldsymbol{q}}|^2 + b_{\text{SC}} \int \textrm{d}^2 \boldsymbol{x} \, |\Delta(\boldsymbol{x})|^4,
\end{equation}
where $\Delta(\boldsymbol{x})$ is the Fourier transform of $\Delta_{\boldsymbol{q}}$. The particle-particle bubble, $\Pi_{\text{SC}}(\boldsymbol{q})$, of the SC obeys $\Pi_{\text{SC}}(\boldsymbol{q}) = \Pi_{\text{SC}}(-\boldsymbol{q})$ and, thus, no diode effect will emerge. Furthermore, the particle-particle bubble will be maximal at $\boldsymbol{q} =0$ and we will focus on the temperature regime where $\Pi_{\text{SC}}(\boldsymbol{q}=0) > 1/g_{\text{sc}}$ where we obtain superconductivity. This will always happen at sufficiently low temperatures due to the logarithmic divergence of $\Pi_{\text{SC}}(\boldsymbol{q}=0)$. 

For the AM, there will also be some finite Cooper-channel interaction with coupling constant $g_{\text{am}}$ and we use a mean-field description,
\begin{equation}
    \mathcal{H}_{\text{AM}} = \sum_{\boldsymbol{k}}  d^\dagger_{\boldsymbol{k}} h_{\boldsymbol{k}}d^{}_{\boldsymbol{k}} + \sum_{\boldsymbol{k},\boldsymbol{q}} \left[ \bar{\Delta}_{\boldsymbol{q}} d^\dagger_{\boldsymbol{k} +\boldsymbol{q}/2} \hat{\Gamma} d^\dagger_{-\boldsymbol{k} +\boldsymbol{q}/2} + \text{H.c.} \right] + \sum_{\boldsymbol{q}} \frac{|\bar{\Delta}_{\boldsymbol{q}}|^2}{g_{\text{am}}}.
\end{equation}
Here, $h_{\boldsymbol{k}}$ is the six band normal-state (two from spin and three from the three sublattices of the kagome lattice) of the AM defined in the main text.
As before, we can obtain the associated free energy,
\begin{equation}
    \mathcal{F}_{\text{AM}} \sim \sum_{\boldsymbol{q}} \left[ \frac{1}{g_{\text{am}}} - \Pi_{\text{AM}}(\boldsymbol{q}) \right] |\bar{\Delta}_{\boldsymbol{q}}|^2 + b_{\text{am}} \int \textrm{d}^2 \boldsymbol{x} \, |\bar{\Delta}(\boldsymbol{x})|^4.
\end{equation}
Importantly, in general, $\Pi_{\text{AM}}(\boldsymbol{q}=0)$ will not diverge at $T\rightarrow 0$ as time-reversal symmetry is broken as a result of altermagnetic order. We will choose choose the coupling $g_{\text{am}}$ sufficiently weak such that $\Pi_{\text{AM}}(\boldsymbol{q}=0) < 1/g_{\text{am}}$ for all temperatures that we consider in our work to describe the fact that the AM material is not a superconductor on its own. 

However, superconducting correlations in the AM will develop as a result of the coupling Hamiltonian $\mathcal{H}_c$. On the level of the above free-energy expansion, it induces a term of the form
\begin{equation}
    \mathcal{F}_{c} \sim - \int \textrm{d}^2 \boldsymbol{x} \left[ T(\boldsymbol{x}) \bar{\Delta}^*(\boldsymbol{x}) \Delta(\boldsymbol{x}) + \text{c.c.} \right], \label{Fcoupling1}
\end{equation}
where $T(\boldsymbol{x}) \in \mathbb{C}$ is generally spatially modulated as a result of a possible moiré pattern at the interface of the two materials. 
Transforming \equref{Fcoupling1} to Fourier space, we find,
\begin{equation}
    \mathcal{F}_{c} \sim - \sum_{\boldsymbol{Q}\in \text{RML}} \sum_{\boldsymbol{q}}  \left[ T_{\boldsymbol{Q}} \bar{\Delta}^*_{\boldsymbol{q}+\boldsymbol{Q}} \Delta_{\boldsymbol{q}} + \text{c.c.} \right], \label{Fcoupling2}
\end{equation}
where $T_{\boldsymbol{q}}$ is the Fourier transform of $T(\boldsymbol{x})$ which is only non-zero for momenta taken from the reciprocal lattice of the emergent moiré superlattice, as indicated by $\boldsymbol{Q}\in \text{RML}$.

Making the common single-$\boldsymbol{q}$ approximation for the computation of the critical currents, i.e., $\Delta_{\boldsymbol{q}} = \delta_{\boldsymbol{q},\boldsymbol{q}_0}$ and $\Delta_{\boldsymbol{q}} = \delta_{\boldsymbol{q},\boldsymbol{q}_1}$, and assuming that the $\boldsymbol{Q}=0$ component of $T_{\boldsymbol{Q}}$ dominates, we find $\boldsymbol{q}_0 = \boldsymbol{q}_1$ as the configuration with the lowest free energy. Then the total free energy at fixed $\boldsymbol{q}_0$ effectively becomes
\begin{equation}
    \mathcal{F}_{\text{eff}} \sim \left[ \frac{1}{g_{\text{sc}}} - \Pi_{\text{SC}}(\boldsymbol{q}_0) \right] |\Delta_{\boldsymbol{q}_0}|^2 + \left[ \frac{1}{g_{\text{am}}} - \Pi_{\text{AM}}(\boldsymbol{q}_0) \right] |\bar{\Delta}_{\boldsymbol{q}_0}|^2 + b_{\text{SC}}  |\Delta_{\boldsymbol{q}_0}|^4 + b_{\text{AM}}  |\bar{\Delta}_{\boldsymbol{q}_0}|^4 - 2 \,\text{Re}[T_{\boldsymbol{Q}=0}\Delta^*_{\boldsymbol{q}_0} \bar{\Delta}^{\phantom{*}}_{\boldsymbol{q}_0} ].
\end{equation}
This is the free energy we use to compute the critical current. More specifically, let us denote the configuration with the lowest energy of $\mathcal{F}_{\text{eff}}$ at given $\boldsymbol{q}_0$ by $(\Delta^{(0)}_{\boldsymbol{q}_0},\bar{\Delta}^{(0)}_{\boldsymbol{q}_0})$. Then the current simply reads as
\begin{equation}
    \boldsymbol{J}_{\boldsymbol{q}} = 2e \left[ - (\partial_{\boldsymbol{q}} \Pi_{\text{SC}}(\boldsymbol{q}) ) |\Delta^{(0)}_{\boldsymbol{q}}|^2  - (\partial_{\boldsymbol{q}} \Pi_{\text{AM}}(\boldsymbol{q}) ) |\bar{\Delta}^{(0)}_{\boldsymbol{q}}|^2 \right].
\end{equation}

\section{Minimizing the effective free energy}
In this section, we provide more details on our procedure for minimizing the free energy.
\\
\\
As a \textit{first step}, we write the effective free energy of our system as, 
\begin{equation}
\mathcal{F}_{\text{eff}}
= 
g_{\text{SC}}(\boldsymbol{q})|\Delta|^{2}
+
g_{\text{AM}}(\boldsymbol{q})|\bar{\Delta}|^{2}
-
\left[
T_{\boldsymbol{Q}=0}
\Delta^{*}
\bar\Delta 
+
T_{\boldsymbol{Q}=0}^{*}\Delta\bar\Delta^{*}
\right]
+
b_{\text{SC}}|\Delta|^{4}
+
b_{\text{AM}}|\bar{\Delta}|^{4} ,
\end{equation}
where we have defined the coefficients,
\begin{equation}
\alpha_{\text{SC}}(\boldsymbol{q}) 
= 
g_{\text{SC}}^{-1} 
-
\Pi_{\text{SC}}(\boldsymbol{q}),
\quad
\alpha_{\text{AM}}(\boldsymbol{q}) 
=
g_{\text{AM}}^{-1} 
-
\Pi_{\text{AM}}(\boldsymbol{q}) .
\end{equation}
As a \textit{second step}, we determine the configuration with lowest effective free energy through the requirements, 
\begin{equation}
\begin{split}
\frac{
\partial \mathcal F_{\text{eff}}
}{
\partial \Delta^*
}
&=
\alpha_{\text{SC}}(\boldsymbol{q}) \Delta 
-
T_{\boldsymbol{Q}=0}\bar\Delta 
+
2b_{\text{SC}}|\Delta|^2\Delta 
= 0
,
\\
\frac{
\partial \mathcal F_{\text{eff}}
}{
\partial \bar\Delta^*
}
&= 
\alpha_{\text{AM}}(\boldsymbol{q}) \bar\Delta 
-
T_{\boldsymbol{Q}=0}^*\Delta 
+
2b_{\text{AM}}|\bar\Delta|^2\bar\Delta 
= 0
\end{split}    
\end{equation}
We now write the superconducting order parameters as $\Delta=\rho e^{i\phi}$
and $\bar\Delta=\bar\rho e^{i\bar\phi}$ with $\rho,\bar\rho\geq0$. Moreover, we parametrize the coupling as
$T_{\boldsymbol{Q}=0}=\tau e^{i\theta}$ with $\tau\geq0$. With these definitions, we can formulate the above requirements as,
\begin{equation}
\begin{split}
& \alpha_{\text{SC}} \rho - \tau\bar\rho e^{i(\theta+\bar\phi-\phi)} + 2b_{\text{SC}}\rho^3 = 0   
\\
&
\alpha_{\text{AM}}\bar\rho - \tau\rho e^{-i(\theta+\bar\phi-\phi)} + 2b_{\text{AM}}\bar\rho^3 = 0.
\end{split}   
\end{equation}
If we take the imaginary part of both equations, we find that, 
\begin{equation}
\begin{split}
\tau\bar\rho\sin(\theta+\bar\phi-\phi)=0,\quad \tau\rho\sin(\theta+\bar\phi-\phi)=0.    
\end{split}    
\end{equation}
Assuming that $\tau,\rho,\bar\rho\neq 0$, these conditions imply that $\sin(\theta+\bar\phi-\phi)=0$ and, hence, 
$
\theta+\bar\phi-\phi = 0$ or $\theta+\bar\phi-\phi = \pi. 
$
The solution with lower free energy is given by $\theta+\bar\phi-\phi = 0$, since the effective free energy contains the term 
$
-(T_{\boldsymbol{Q}=0}
\Delta^{*}
\bar\Delta 
+
T_{\boldsymbol{Q}=0}^{*}\Delta\bar\Delta^{*})
=
-2\tau\rho\bar\rho\cos(\theta+\bar\phi-\phi)
$. Consequently, the requirements for minimizing the free energy can be written more compactly as, 
\begin{equation}
\begin{split}
& \alpha_{\text{SC}}(\boldsymbol{q})  \rho - \tau\bar\rho+ 2b_{\text{SC}}\rho^3 = 0   
\\
&
\alpha_{\text{AM}}(\boldsymbol{q}) \bar\rho - \tau\rho + 2b_{\text{AM}}\bar\rho^3 = 0.
\end{split}   
\end{equation}
As a \textit{third step}, we want to determine a solution to these requirements for minimizing the effective free energy. For this purpose, we assume that the altermagnet never condenses by itself, $\alpha_{\text{AM}}(\boldsymbol{q}) >0$, and that the induced superconducting order parameter, $\bar\Delta$, is sufficiently small. Under these conditions the term $2b_{\text{AM}}\bar\rho^3$ is a negligible correction to the second requirement. Hence, we have, 
\begin{equation}
\bar\rho \approx \frac{\tau}{\alpha_{\text{AM}}(\boldsymbol{q}) }\rho.    
\end{equation}
If we insert this condition into the first requirement, we find that, 
\begin{equation}
\left(
\alpha_{\text{SC}}(\boldsymbol{q}) 
-
\frac{\tau^2}{\alpha_{\text{AM}}(\boldsymbol{q}) }
\right)
\rho + 2b_{\text{SC}}\rho^3 = 0    
\end{equation}
We now define an effective coefficent, 
\begin{equation}
 \alpha^{\text{eff}}_{\text{SC}}(\boldsymbol{q})
 =
 \alpha_{\text{SC}}(\boldsymbol{q}) 
-
\frac{\tau^2}{\alpha_{\text{AM}}(\boldsymbol{q}) }   
=
\left( 
g_{\text{SC}}^{-1} 
-
\Pi_{\text{SC}}(\boldsymbol{q})
\right)
-
\frac{|T_{\boldsymbol{Q}=0}|^{2}
}
{
g_{\text{AM}}^{-1} 
-
\Pi_{\text{AM}}(\boldsymbol{q})
} 
\end{equation}
We assume that the effective coupling is sufficiently weak so that 
$\alpha^{\text{eff}}_{\text{SC}}(\boldsymbol{q})\leq 0$ since $\alpha_{\text{SC}}(\boldsymbol{q}) \leq 0$. 
We then find the following expressions for the renormalized gap in the superconductor 
and the induced gap, 
\begin{equation}
\rho^2 = -\frac{\alpha^{\text{eff}}_{\text{SC}}(\boldsymbol{q})}{2 b_{\text{SC}}},
\quad
\bar{\rho}^2 = \left( \frac{\tau}{\alpha_{\text{AM}}(\boldsymbol{q}) } \right)^2 \rho^2 . 
\end{equation}
We can also write the expressions, equivalently, in the form, 
\begin{equation}
|\Delta^{(0)}_{\boldsymbol{q}}|^{2}
=
-
\frac{
\alpha^{\text{eff}}_{\text{SC}}(\boldsymbol{q})
}
{
2 b_{\text{SC}}
},
\quad
|\bar{\Delta}^{(0)}_{\boldsymbol{q}}|^{2}
=
\left(
\frac{
|T_{\boldsymbol{Q}=0}|
}
{
g_{\text{AM}}^{-1} - \Pi_{\text{AM}}(\boldsymbol{q})
}
\right)^{2}
|\Delta^{(0)}_{\boldsymbol{q}}|^{2}.
\end{equation}
This concludes our derivation of the order parameters that minimize the effective free energy.

\section{Evaluation of the superconducting order parameters} 
In this section, we provide additional details on the evaluation of the superconducting order parameters. 

\subsection{Proximity-induced superconducting order parameter}
We first focus on the proximity-induced superconducting order parameter,
\begin{equation}
|\bar{\Delta}^{(0)}_{\boldsymbol{q}}|^{2}
=
\left(
\frac{
|T_{\boldsymbol{Q}=0}|
}
{
g_{\text{AM}}^{-1} - \Pi_{\text{AM}}(\boldsymbol{q})
}
\right)^{2}
|\Delta^{(0)}_{\boldsymbol{q}}|^{2}.    
\end{equation}
In the weak-coupling regime, we want to ensure that the proximity-induced order parameter in the altermagnet never exceeds the order parameter in the parent superconductor. We, therefore, require that
the ratio of the superconducting order parameters is bounded, 
\begin{equation}
\frac{|\bar\Delta^{(0)}_{\boldsymbol q}|}{|\Delta^{(0)}_{\boldsymbol q}|}
=
\left|
\frac{T_{\boldsymbol{Q}=0}}{\alpha_{\text{AM}}(\boldsymbol{q}) }
\right|
\leq
r_{\max}
\end{equation}
for all $\boldsymbol{q}$. Here, $r_{\max}$ is the maximum ratio with $0 < r_{\max} \leq 1$. 
\\
\\
To ensure that the above inequality holds for all $\boldsymbol{q}$, we recall that $\alpha_{\text{AM}}(\boldsymbol{q}) >0$ since the altermagnet never condenses. In particular, $\alpha_{\text{AM}}(\boldsymbol{q}) $ is bounded from below,
\begin{equation}
\alpha_{\text{AM}}(\boldsymbol{q})  \geq \alpha_{\text{AM,min}}  \equiv \frac{1}{g_{\text{AM}}}-\Pi_{\text{AM}}(\boldsymbol{q}_{\text{max}}),
\end{equation}
where $\boldsymbol{q}_{\text{max}}$ corresponds to the momentum where $\Pi_{\text{AM}}(\boldsymbol{q})$
takes on its maximum value. At the same time, we also know that, 
\begin{equation}
\alpha_{\text{AM,min}} \geq \frac{|T_{\boldsymbol{Q}=0}|}{r_{\max}}    
\end{equation}
since 
\begin{equation}
 \left|
\frac{T_{\boldsymbol{Q}=0}}{\alpha_{\text{AM,min}} }
\right|
\leq
r_{\max}   
\end{equation}
As a result, we arrive at the following condition,
\begin{equation}
\frac{1}{g_{\text{AM}}}
\geq
\Pi_{\text{AM}}(\boldsymbol{q}_{\text{max}})
+
\frac{|T_{\boldsymbol{Q}=0}|}{r_{\max}},
\end{equation}
which sets a lower bound on $1/g_{\text{AM}}$.

\subsection{Parent superconducting order parameter}
The superconducting order parameter in the parent superconductor is renormalized due to the altermagnet and given by, 
\begin{equation}
|\Delta^{(0)}_{\boldsymbol{q}}|^{2}
=
-
\frac{
\alpha^{\text{eff}}_{\text{SC}}(\boldsymbol{q})
}
{
2 b_{\text{SC}}
}    
=
-
\frac{
1
}
{
2 b_{\text{SC}}
}
\left[
\left( 
g_{\text{SC}}^{-1} 
-
\Pi_{\text{SC}}(\boldsymbol{q})
\right)
-
\frac{|T_{\boldsymbol{Q}=0}|^{2}}{\alpha_{\text{AM}}(\boldsymbol{q}) }   
\right]
\end{equation}
Here, we choose $g_{\text{SC}}$ such that $g_{\text{SC}}^{-1} 
-
\Pi_{\text{SC}}(\boldsymbol{q}=0)<0$. Since $\Pi_{\text{SC}}(\boldsymbol{q}=0)$ is the maximum of the particle-particle bubble for the conventional superconductor, this choice ensures that $g_{\text{SC}}^{-1} 
-
\Pi_{\text{SC}}(\boldsymbol{q})<0$ for any $\boldsymbol{q}$. Moreover, since $\alpha_{\text{AM}}(\boldsymbol{q}) >0$, we have  $\alpha^{\text{eff}}_{\text{SC}}(\boldsymbol{q})<0$ and hence the superconducting order parameter is always well-defined.

\section{Particle-particle bubble of the altermagnet}
In this section, we derive an expression for evaluating the particle-particle bubble of the altermagnet.
\\
\\
As a \textit{first step}, we recall that the particle-particle bubble was defined as, 
\begin{equation}
\Pi_{\text{AM}}(\boldsymbol{q}) 
=
T\sum_{\omega_{n}}\sum_{\boldsymbol{k}}
\operatorname{Tr}\left[
\hat\Gamma_{\text{AM}}
\hat G_{\text{AM}}(\boldsymbol{k}_{+},i\omega_{n})
\hat\Gamma_{\text{AM}}^{T}
\hat G_{\text{AM}}^{T}(-\boldsymbol{k}_{-}, -i\omega_{n})
\right]
\end{equation}
with $\boldsymbol{k}_{\pm}=\boldsymbol{k}\pm \frac{\boldsymbol{q}}{2}$. Here, the normal-state Green's function is given by 
$
\hat G_{\text{AM}}(\boldsymbol{k},i\omega_{n})
=
(i\omega_{n}-\hat\xi^{(\text{AM})}_{\boldsymbol{k}})^{-1}
$
with 
$
\hat\xi^{(\text{AM})}_{\boldsymbol{k}}\equiv \hat{h}^{(\text{AM})}_{\boldsymbol{k}}-\mu \mathbb{I}_{6}.
$
Moreover, the pairing vertex is given by 
$
\hat\Gamma_{\text{AM}} = i s_{y}\otimes \mathbb{I}_{3}
$,
which satisfies 
$\hat\Gamma^{T}_{\text{AM}}=-\hat\Gamma_{\text{AM}}$ and $\hat\Gamma_{\text{AM}}^{\dagger}=-\hat\Gamma_{\text{AM}}$.
\\
\\
As a \textit{second step}, we move to a diagonal representation of our Hamiltonian via, 
\begin{equation}
\hat{h}^{(\text{AM})}_{\boldsymbol{k}}
=
U_{\boldsymbol{k}}
\text{diag}
\left(
\varepsilon^{(\text{AM})}_{\boldsymbol{k}1},...,\varepsilon^{(\text{AM})}_{\boldsymbol{k}6}\right)
U_{\boldsymbol{k}}^{\dagger},  
\end{equation}
or, equivalently, in terms of the normal state Green's function, 
\begin{equation}
\begin{split}
\hat{G}_{\text{AM}}(\boldsymbol{k},i\omega_{n})
&=
U_{\boldsymbol{k}}
D_{\text{AM}}(\boldsymbol{k},i\omega_{n})
U_{\boldsymbol{k}}^{\dagger}    
\\
\hat G_{\text{AM}}^{T}(\boldsymbol{k},i\omega_{n})
&=
(
U_{\boldsymbol{k}} D_{\text{AM}} U_{\boldsymbol{k}}^{\dagger}
)^{T}
=
U_{\boldsymbol{k}}^{*}
D_{\text{AM}}(\boldsymbol{k},i\omega_{n}) 
U_{\boldsymbol{k}}^{T} 
\quad
\text{with}
\quad
D_{\text{AM}}(\boldsymbol{k},i\omega_{n})
=
\text{diag}
\left(
\frac{1}{i\omega_{n}-\xi_{\boldsymbol{k}1}}
,...,
\frac{1}{i\omega_{n}-\xi_{\boldsymbol{k}6}}
\right),
\end{split}    
\end{equation}
and $\xi^{(\text{AM})}_{\boldsymbol{k}\nu} \equiv \varepsilon^{(\text{AM})}_{\boldsymbol{k}\nu} - \mu$.
\\
\\
As a \textit{third step}, we insert these results into the expressions for the particle-particle bubble, 
\begin{equation}
\Pi_{\text{AM}}(\boldsymbol{q})
=
T\sum_{\omega_{n}}\sum_{\boldsymbol{k}}
\text{Tr}
\left[
\hat\Gamma_{\text{AM}}
U_{\boldsymbol{k}_{+}} 
D_{\text{AM}}(\boldsymbol{k}_{+},i\omega_{n})
U_{\boldsymbol{k}_{+}}^{\dagger}
\hat\Gamma_{\text{AM}}^{T}
U_{-\boldsymbol{k}_{-}}^{*} 
D_{\text{AM}}(-\boldsymbol{k}_{-},-i\omega_{n})
U_{-\boldsymbol{k}_{-}}^{T}
\right].
\end{equation}
We can now use the cyclic property of the trace to rewrite this expression as, 
\begin{equation}
\Pi_{\text{AM}}(\boldsymbol{q})
=
T\sum_{\omega_{n}}\sum_{\boldsymbol{k}}
\text{Tr}
\left[
U_{-\boldsymbol{k}_{-}}^{T} 
\hat\Gamma_{\text{AM}} 
U_{\boldsymbol{k}_{+}}
D_{\text{AM}}(\boldsymbol{k}_{+},i\omega_{n})
U_{\boldsymbol{k}_{+}}^{\dagger} 
\hat\Gamma^{T}_{\text{AM}} 
U_{-\boldsymbol{k}_{-}}^{*}
D_{\text{AM}}(-\boldsymbol{k}_{-},-i\omega_{n})
\right].
\end{equation}
To further simplify this expression, we introduce the rotated pairing vertex, 
\begin{equation}
\begin{split}
M^{(\text{AM})}(\boldsymbol{k},\boldsymbol{q})
&=
U_{-\boldsymbol{k}_{-}}^{T}
\hat\Gamma_{\text{AM}} 
U_{\boldsymbol{k}_{+}}
\\
(M^{(\text{AM})}(\boldsymbol{k},\boldsymbol{q}))^{\dagger}
&=
(
U_{-\boldsymbol{k}_{-}}^{T} 
\hat\Gamma
U_{\boldsymbol{k}_{+}}
)^{\dagger}
=
U_{\boldsymbol{k}_{+}}^{\dagger} 
\hat\Gamma^{T} 
U_{-\boldsymbol{k}_{-}}^{*} 
\end{split}
\end{equation}
Here, we have used that $\hat\Gamma^{T}=-\hat\Gamma$ and $\hat\Gamma^{\dagger}=-\hat\Gamma$. With this definition, the particle-particle bubble takes on the form, 
\begin{equation}
\Pi_{\text{AM}}(\boldsymbol{q})
=
T\sum_{\omega_{n}}\sum_{\boldsymbol{k}}
\text{Tr}\left[
M^{(\text{AM})}(\boldsymbol{k},\boldsymbol{q})
D_{\text{AM}}(\boldsymbol{k}_{+},i\omega_{n})
(M^{(\text{AM})}(\boldsymbol{k},\boldsymbol{q}))^{\dagger}
D_{\text{AM}}(-\boldsymbol{k}_{-},-i\omega_{n})
\right]. 
\end{equation}
\\
\\
As a \textit{fourth step}, we want to explicitly evaluate the trace in our expression for the particle-particle bubble. For this purpose, we write the matrix products explicitly as, 
\begin{equation}
\begin{split}
\Pi_{\text{AM}}(\boldsymbol{q})
&=
T
\sum_{\omega_{n}}\sum_{\boldsymbol{k}}
\sum_{\mu,\nu} 
M^{(\text{AM})}(\boldsymbol{k},\boldsymbol{q})_{\mu\nu}
\frac{
1
}{
i\omega_{n}-\xi_{\boldsymbol{k}_{+},\nu}
}
(
M^{(\text{AM})}(\boldsymbol{k},\boldsymbol{q})^{\dagger}
)_{\nu\mu}
\frac{
1
}{
-i\omega_{n}-\xi_{-\boldsymbol{k}_{-},\mu}
}
\\
&=
T\sum_{\omega_{n}}
\sum_{\boldsymbol{k}}
\sum_{\mu,\nu}
\frac{
\left|
M^{(\text{AM})}_{\mu\nu}\left(\boldsymbol{k},\boldsymbol{q}\right)
\right|^{2}
}
{
\left(
i\omega_{n}-\xi^{(\text{AM})}_{\boldsymbol{k}_{+},\nu}
\right)
\left(
-i\omega_{n}-\xi^{(\text{AM})}_{-\boldsymbol{k}_{-},\mu}
\right)
}
\end{split}
\end{equation}
where $\mu,\nu=1,...,6$. 
\\
\\
As a \textit{fifth step}, we remark that the sum over Matsubara frequencies can be explicitly evaluated,  
\begin{equation}
\Pi_{\text{AM}}(\boldsymbol{q})
=
\sum_{\boldsymbol{k}}
\sum_{\mu,\nu}
\left|
M^{(\text{AM})}_{\mu\nu}(\boldsymbol{k},\boldsymbol{q})
\right|^{2}
\frac{
1-f(\xi^{(\text{AM})}_{\boldsymbol{k}_{+},\nu
})
-f(\xi^{(\text{AM})}_{-\boldsymbol{k}_{-},\mu})
}
{
\xi^{(\text{AM})}_{\boldsymbol{k}_{+},\nu}+\xi^{(\text{AM})}_{-\boldsymbol{k}_{-},\mu}
}
=
\sum_{\boldsymbol{k}} 
\sum_{\mu,\nu}
\left| 
M^{(\text{AM})}_{\mu\nu} 
\right|^{2}
\frac{
\tanh\left(
\frac{\xi^{(\text{AM})}_{\boldsymbol{k}_{+},\nu}}{2T}
\right)
+ 
\tanh\left( 
\frac{\xi^{(\text{AM})}_{-\boldsymbol{k}_{-},\mu}}{2T} 
\right)
}{
2( 
\xi^{(\text{AM})}_{\boldsymbol{k}_{+},\nu} 
+
\xi^{(\text{AM})}_{-\boldsymbol{k}_{-},\mu} 
)
}
.    
\end{equation}
For numerical evaluations, it is useful to note that this expression is well-behaved in the limit when 
$\xi^{(\text{AM})}_{\boldsymbol{k}_{+},\nu} + \xi^{(\text{AM})}_{-\boldsymbol{k}_{-},\mu} \rightarrow 0$. Specifically, in this case, 
\begin{equation}
\frac{
1 - f(\xi^{(\text{AM})}_{\boldsymbol{k}_{+},\nu})
-
f(\xi^{(\text{AM})}_{-\boldsymbol{k}_{-},\mu})
}
{
\xi^{(\text{AM})}_{\boldsymbol{k}_{+},\nu} 
+
\xi^{(\text{AM})}_{-\boldsymbol{k}_{-},\mu}
}
\rightarrow
\frac{1}{4T}\text{sech}^{2}
\left(
\frac{
\xi^{(\text{AM})}_{\boldsymbol{k}_{+},\nu}
}
{
2T
}
\right).
\end{equation}
This completes our derivation of the particle-particle bubble for the altermagnet.

\end{document}